\documentclass[prb,reprint,amssymb,showpacs,aps]{revtex4-1}
\usepackage{graphicx}
\usepackage{amsmath}
\usepackage{braket}
\usepackage{subfigure}
\usepackage{bm}

\usepackage{overpic}
\usepackage{tikz}

\begin{document}

\title{Renormalization Group Approach for the Wave Packet Dynamics in Golden-Mean and Silver-Mean Labyrinth Tilings}

\author{Stefanie Thiem}
\author{Michael Schreiber}
\affiliation{Institut f\"ur Physik, Technische Universit\"at Chemnitz, D-09107 Chemnitz, Germany}
\date{\today}

\begin{abstract}
We study the quantum diffusion in quasiperiodic tight-binding models in one, two, and three dimensions. First, we investigate a class of one-dimensional quasiperiodic chains, in which the atoms are coupled by weak and strong bonds aligned according to the metallic-mean sequences. The associated generalized labyrinth tilings in $d$ dimensions are then constructed from the direct product of $d$ such chains, which allows us to consider rather large systems numerically. The electronic transport is studied by computing the scaling behavior of the mean square displacement of the wave packets with respect to time. The results reveal the occurrence of anomalous diffusion in these systems. By extending a renormalization group approach, originally proposed for the golden-mean chain, we show also for the silver-mean chain as well as for the higher-dimensional labyrinth tilings that in the regime of strong quasiperiodic modulation the wave-packet dynamics are governed by the underlying quasiperiodic structure.
\end{abstract}


\maketitle

\section{Introduction}\label{sec:introduction}

Understanding the relations between the atomic structure and the physical properties of materials remains one of the elementary questions of condensed-matter physics. One research line in this quest started with the discovery of quasicrystals by Shechtman et al. in 1982.\cite{PhysRevLett.1984.Shechtman} It soon became clear that quasicrystals with their 5-, 8-, 10- or 12-fold rotational symmetries, forbidden according to classical crystallography, can be described in terms of mathematical models of aperiodic tilings of a plane proposed by Penrose and Ammann in the 1970s.\cite{DCGeom.1992.Ammann, BIMathApp.1974.Penrose} Further, quasicrystals are often regarded to have a degree of order intermediate between crystals and disordered systems, which can be understood by considering the repetitiveness of local patterns. While any local pattern in crystals repeats for integer combinations of the lattice vectors, Conway's theorem states that in a quasicrystal a given local pattern in a region of some diameter $L$ will be repeated within a distance of two diameters $2L$.\cite{SciAm.1977.Gardner} In liquids or amorphous systems the distance of identical local patterns of size $L$ grows exponentially with $L$. Moreover, the deterministic construction rules of the quasicrystal models can also be interpreted as some kind of order in contrast to the situation in disordered systems. On the other hand, translational symmetry is missing in contrast to crystals.

Various experimental investigations revealed rather exotic electrical, magnetic, and optical properties of these materials.\cite{UsefulQuasicrystals, PhysicalProperties.1999.Stadnik} As many quasicrystals contain a high percentage of well-conducting elements, it was originally assumed that their thermal and electrical transport properties were similar to those in crystalline or amorphous metals.\cite{UsefulQuasicrystals} However, it turned out that they rather behave like semiconducting or insulating materials because they often possess a large resistivity for small temperatures and the resistivity decreases with increasing temperature and increasing structural order of the materials.\cite{JMathPhys.1997.Roche, PhysicalProperties.1999.Stadnik, PhysRevLett.1993.Mayou, AdvPhys.1992.Poon}

This motivated extensive research to get a better theoretical understanding of the structure and physical properties of quasicrystals. Today several exact results are known for one-dimensional quasiperiodic systems.\cite{MathQuasi.2000.Damanik, PhysRevB.1987.Kohmoto, JPhysFrance.1989.Sire} However, the properties in two or three dimensions have been clarified to much lesser degree and are mainly based on numerical studies for limited systems with a few thousands sites due to the missing translational symmetry of quasicrystals. For instance, the electronic properties were determined numerically for the Penrose tiling, the octagonal tiling, and the Ammann-Kramer-Neri tiling.\cite{Quasicrystals.2003.Grimm, PhysRevB.1992.Passaro} However, this makes it difficult to understand the nature of eigenstates and properties of macroscopic systems. To address this challenge, different approaches have been followed. For example, by expanding the time-evolution operator in Chebyshev polynomials the spreading of energy-filtered wave packets was studied for generalized Rauzy tilings with up to $10^6$ sites.\cite{PhysRevB.2002.Triozon, PhysRevB.2003.Vidal} However, this approach only works for systems with a rather smooth local density of states. Further, it is possible to calculate some exact eigenstates of the tight-binding Hamiltonian on the Penrose tiling. \cite{PhysRevB.1995.Rieth,PhysRevB.1998.Repetowicz}

We follow another approach by studying the wave-packet dynamics for $d$-dimensional quasicrystalline models with a separable Hamiltonian in a tight-binding approach.\cite{EurophysLett.1989.Sire} This method is based on mathematical sequences, constructed by an inflation rule $\mathcal{P}$ describing the weak and strong couplings of atoms in a quasiperiodic chain. The higher-dimensional labyrinth tilings are then constructed as a direct product of these chains and their eigenstates are directly calculated by multiplying the energies or wave functions of these chains. This allows us to study very large systems in higher dimensions with up to $10^{10}$ sites based on the solutions in one dimension. However, the property of separability comes at the price that the labyrinth tilings do not possess a non-crystallographic rotational symmetry. Nevertheless, the labyrinth tiling is able to describe a subset of points of the octagonal quasicrystal associated to the structure of actual quasicrystals with 8-fold rotational symmetry.\cite{JPhysFrance.1989.Sire2,EurophysLett.1989.Sire}

In this paper we relate the hierarchical properties of these quasiperiodic systems to their electronic transport properties by numerical calculations and a renormalization group (RG) approach in the regime of strong quasiperiodic modulation of the bond strength. Up to now this method has only been used to describe the quantum diffusion for the one-dimensional Fibonacci chain,\cite{PhysRevA.1987.Abe,JPhysJap.1988.Hiramoto, PhysRevLett.1996.Piechon} whereas in this paper we will also study the wave-packet dynamics for the silver-mean chain and the generalized labyrinth tilings.

The upcoming sections are structured as follows: In Sec.\ \ref{sec:model} we introduce the metallic-mean chains and describe the construction of the higher-dimensional labyrinth tilings. In Sec.\ \ref{sec:num-results} we then present numerical results for wave-packet dynamics in these quasiperiodic tilings.
Using an RG approach and perturbation theory we derive also analytical results for the scaling behavior of the width of the wave packet in Secs.\ \ref{sec:RG-theory} to \ref{sec:rg-labyrinth}, which show a good agreement with the numerical results in the regime of strong quasiperiodic modulation. The results are briefly summarized in Sec.\ \ref{sec:conclusion}.

\section{Generalized Labyrinth Tilings}
\label{sec:model}

The construction of the $d$-dimensional generalized labyrinth tilings is based on $d$ so-called metallic-mean quasiperiodic sequences, which for a parameter $b$ are defined by the inflation rule\cite{PhysRevB.2009.Thiem}
\begin{equation}
 \label{equ:octonacci.1} \mathcal{P}_b =
 \begin{cases}
   s \longrightarrow w \\
   w \longrightarrow wsw^{b-1}
 \end{cases}\quad.
\end{equation}
Starting with the symbol $s$ we obtain after $a$ iterations the $a$th order approximant $\mathcal{C}_a$ of the quasiperiodic chain. The length $f_a$ of an approximant $\mathcal{C}_a$ is given by the recursive rule $f_a = b f_{a-1} + f_{a-2}$ with $f_0 = f_1 = 1$. Further, the ratio of the lengths of two successive iterants as well as the ratio of the numbers $\#$ of occurrence of the symbols $w$ and $s$ in an approximant approach different metallic means for $a \to \infty$.\cite{JPhysA.1989.Gumbs} Thus, depending on the parameter $b$ we obtain with the continued fraction representation $\tau_b = [\bar{b}] = [b,b,b,...]$ the relations
\begin{equation}
\label{equ:tau}
\lim_{a \to \infty} \frac{f_a}{f_{a-1}} = \lim_{a \to \infty} \frac{\#_w\left( \mathcal{C}_a \right) } {\#_s\left( \mathcal{C}_a \right)} =  \tau_b \;.
\end{equation}
In this paper we only consider the cases $b=1$ and $b=2$. Thereby, the first case corresponds to the well known Fibonacci sequence with the golden mean $\tau_{\mathrm{Au}} = [\bar{1}]=(1+\sqrt{5})/2$, which is related to real quasicrystals with 5- and 10-fold symmetry,\cite{PhysRevB.1986.Socolar} and the latter case results in the octonacci sequence with silver mean $\tau_{\mathrm{Ag}} = [\bar{2}] = 1+\sqrt{2}$, which is related to octagonal quasicrystals.\cite{JPhysFrance.1989.Sire2}

Our model describes an electron hopping from one vertex of a quasiperiodic chain to a neighboring one. The aperiodicity is given by the underlying quasiperiodic sequence of couplings, where the symbols $w$ and $s$ denote the weak and strong bonds in the chain. Solving the corresponding time-independent Schr\"odinger equation
\begin{align}
 \label{equ:octonacci.8}
 \mathbf{H} \ket{\Psi^i}  =  E^i \ket{\Psi^i} \Leftrightarrow t_{l-1,l} \Psi_{l-1}^i + t_{l,l+1} \Psi_{l+1}^{i} =  E^i \Psi_l^i
 \end{align}
for the quasiperiodic systems with zero on-site potentials, we obtain discrete energy values $E^i$ and wave functions $\ket{\Psi^i} = \sum_{l=1}^{f_a+1} \Psi_l^i \ket{l}$ represented in the orthogonal basis states $\ket{l}$ associated to a vertex $l$. The hopping strength $t$ in the Schr\"odinger equation is given by the quasiperiodic sequence $\mathcal{C}_a$ with $t_{s} = s$ for a strong bond and $t_{w} = w$ for a weak bond ($0 < w \le s$).\cite{Quasicrystals.2003.Grimm,PhysRevB.2000.Yuan} Applying free boundary conditions the number of vertices is $N_a = f_a + 1$.

\begin{figure}[t!]
 \includegraphics[width=0.8\columnwidth]{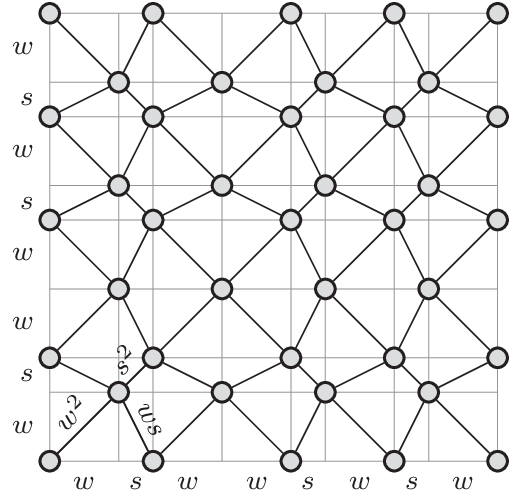}
 \caption{Golden-mean labyrinth tiling $\mathcal{L}_5^\mathrm{Au}$ constructed from two golden-mean chains $\mathcal{C}_5^\mathrm{Au}$, which are perpendicular to each other. The bond strengths of the labyrinth tiling equal the product of the associated bond strengths of the one-dimensional chains as denoted for the three different bond types in the left lower corner.}
  \label{fig:labyrinth}
\end{figure}

The results show that the eigenvalues are symmetric with respect to $0$. Thus, for even system sizes $N_a$ all energies $E$ have a symmetric counterpart $-E$, but for odd $N_a$ there is one state $E = 0$, which has no corresponding state. Additionally, the eigenfunctions possess a symmetry:\cite{PhysRevB.2000.Yuan} The eigenstate $\Psi$ with the eigenvalue $E$ and the eigenstate $\widetilde{\Psi}$ with the respective eigenvalue $-E$ only differ by an alternating sign depending on the vertex $l$ according to $\widetilde{\Psi}_l = (-1)^l \Psi_l$. Further, for odd $N_a$ the eigenvector $\Psi_l$ associated to the eigenvalue $E = 0$ has a special structure, i.e., $\Psi_l$ vanishes on all even sites $l$.\cite{PhysRevB.2000.Yuan}

In a next step the generalized labyrinth tilings $\mathcal{L}_a^{d\mathrm{d}}$ in $d$ dimensions are constructed from the Euclidian product $\mathcal{C}_a \times \mathcal{C}_a \times \ldots \times \mathcal{C}_a$ of $d$ quasiperiodic chains $\mathcal{C}_a$.\cite{EurophysLett.1989.Sire, PhysRevB.2005.Cerovski} From a geometrical point of view these chains are perpendicular to each other and the diagonals of the resulting grid correspond to the bonds of the labyrinth tiling as shown in Fig.\ \ref{fig:labyrinth}. Due to this product approach also the bond strengths of the labyrinth tiling equal the products of the corresponding bond strengths of the one-dimensional chains as visualized in Fig.\ \ref{fig:labyrinth}. Hence, in $d$ dimension $d+1$ different couplings occur.
Note that the grid decomposes into $2^{d-1}$ separate grids depending on the starting point. Each of these grids corresponds to a finite $a$th order approximant $\mathcal{L}_a$ of the generalized labyrinth tiling $\mathcal{L}$.\cite{JPhysFrance.1989.Sire2,PhysRevB.2000.Yuan}

The Hamiltonian of the higher-dimensional generalized labyrinth tiling is separable. Thus, the eigenstates of the labyrinth in $d$ dimensions can be constructed from the product of the eigenstates of $d$ one-dimensional chains:\cite{PhysRevB.2000.Yuan}\begin{subequations}
\label{equ:eigenstates}
\begin{align}
    E^\mathbf{s} &= E^{ij\ldots k} = E^{1i} E^{2j}\ldots  E^{dk} \\
    \Phi_\mathbf{r}^\mathbf{s} &= \Phi_{lm\ldots n}^{ij\ldots k} \propto \Psi_{l}^{1i} \Psi_{m}^{2j} \ldots \Psi_{n}^{dl} \;.
\end{align}
\end{subequations}
The superscripts $\mathbf{s} = (i,j,\ldots,k)$ enumerate the eigenvalues $E$ and $\mathbf{r}=(l,m,\ldots,n)$ represent the coordinates of the vertices in the quasiperiodic tiling.
Due to the symmetries of the eigenstates of the one-dimensional chains mentioned above, some of the wave functions $\Phi_\mathbf{r}^\mathbf{s}$ and related eigenvalues $E^\mathbf{s}$ are identical and we have to choose a subset which only includes distinct eigenstates. The construction of such a subset is described in detail elsewhere.\cite{PhysRevB.2000.Yuan, JPhysCS.2010.Thiem}

\section{Mean-Square Displacement of a Wave Packet}
\label{sec:num-results}

\begin{figure*}[t!]
  \centering
  \includegraphics[width=0.33\textwidth]{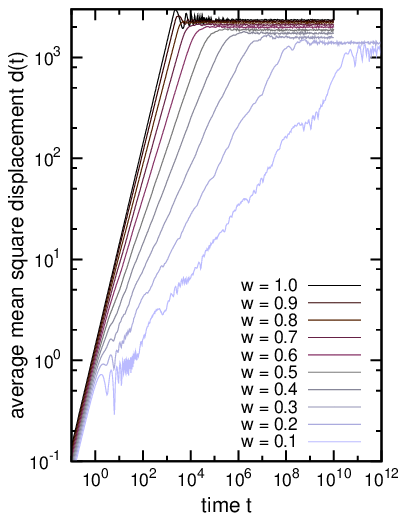}\hfill
  \includegraphics[width=0.33\textwidth]{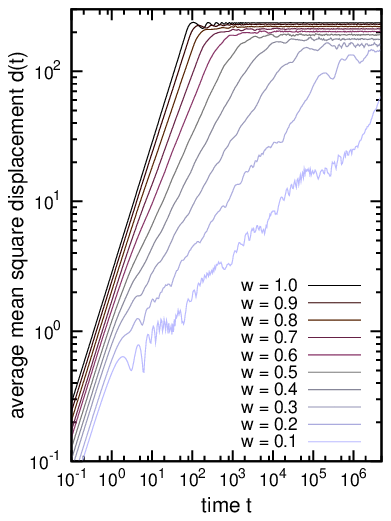}\hfill
  \includegraphics[width=0.33\textwidth]{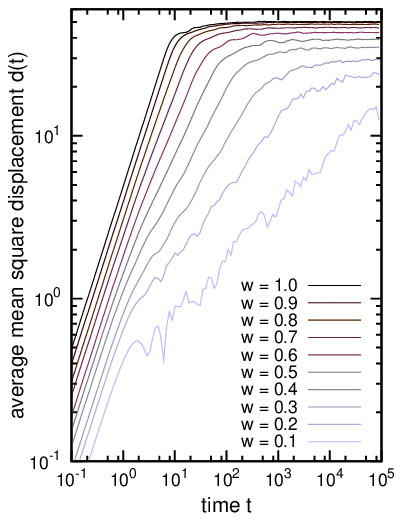}
  \caption{Mean square displacement $d(t)$ of a wave packet averaged over different initial positions for the silver-mean chain $\mathcal{C}_{11}^\textrm{Ag}$ (left), the two-dimensional labyrinth tiling $\mathcal{L}_{8}^\textrm{Ag}$ (center), and the three-dimensional labyrinth tiling $\mathcal{L}_{6}^\textrm{3d,Ag}$ (right) for $s=1$.}
  \label{fig:msd-octonacci}
\end{figure*}

As outlined in the introduction, the connection of the transport properties and the quasiperiodic structure of a system is not yet fully understood, and thus the investigation of transport properties of these materials continues to be of special interest. In this section, we study the wave-packet dynamics in the metallic-mean systems, in particular in the limit of weak coupling ($w \ll s$). The results are then related to the structure of the systems by extending an RG approach in the following section.

We investigate the transport properties of the quasiperiodic systems by means of the time evolution of a wave packet $\ket{\Upsilon(\mathbf{r}_0,t)} = \sum_{\mathbf{r} \in \mathcal{L}} \Upsilon_\mathbf{r}(\mathbf{r}_0,t) \ket{\mathbf{r}}$, which is constructed from the solutions $\Phi_\mathbf{r}(t)$ of the time-dependent Schr\"odinger equation $\mathbf{H}(\mathbf{r},t) \Phi_\mathbf{r}(t) = \mathrm{i} \hbar \dot{\Phi}_\mathbf{r}(t)$. The time-dependent wave functions can be easily obtained via the separation approach from the solutions of the time-independent Schr\"odinger equation in Eq.\ \eqref{equ:eigenstates} according to $\Phi_\mathbf{r}^\mathbf{s}(t) = \Phi_\mathbf{r}^\mathbf{s} e^{-\mathrm{i}E^\mathbf{s}t}$. The wave packet is then represented as a superposition of these orthonormal eigenstates $\Phi_\mathbf{r}^\mathbf{s}(t)$,\cite{JPhys.1995.Zhong} where we assume that it is initially localized at the position $\mathbf{r}_0$ of the quasiperiodic tiling, i.e., $\Upsilon_\mathbf{r} (\mathbf{r}_0,t=0) = \delta_{\mathbf{r}\mathbf{r}_0}$. Hence, with the completeness relation and normalized basis states the wave packet is defined by
\begin{equation}
 \label{equ:wave-pacekt}
 \Upsilon_{\mathbf{r}}(\mathbf{r}_0,t)
  = \sum_{\mathbf{s}} \Phi_{\mathbf{r}_0}^{\mathbf{s}} \Phi_{\mathbf{r}}^{\mathbf{s}}  e^{-\mathrm{i} E^{\mathbf{s}} t} \;.
\end{equation}

Besides calculating the expansion of the wave packet in space, a more detailed analysis can be obtained by computing the mean square displacement (also called width)
\begin{equation}
 d(\mathbf{r}_0,t) = \left[\sum_{\mathbf{r} \in \mathcal{L}} |\mathbf{r}-\mathbf{r}_0|^2 \, |\Upsilon_\mathbf{r}(t)|^2 \right]^{\frac{1}{2}}
\end{equation}
of the wave packet.
The spreading of the width $d(\mathbf{r}_0,t)$ of the wave packet in an infinite system shows anomalous diffusion for $t \rightarrow \infty$ according to $d(\mathbf{r}_0, t) \propto t^{\beta(\mathbf{r}_0)}$, where the scaling exponent $\beta(\mathbf{r}_0)$ depends on the initial position $\mathbf{r}_0$ of the wave packet.\cite{RevMathPhys.1998.SchulzBaldes, JMathPhys.1997.Roche, AdvPhys.1992.Poon, PhysRevLett.1994.Huckestein, PhysRevB.2000.Yuan} The properties of the electronic transport are governed by the wave-packet dynamics averaged over different initial positions, that is, we determine
 \begin{equation}\index{scaling exponent!of mean square displacement}
  \label{equ:msd-scaling}
   d(t) = \langle d(\mathbf{r}_0,t) \rangle \propto t^{\beta} \;.
 \end{equation}
Thereby, the scaling exponent $\beta$ is related to the conductivity $\sigma$ via the generalized Drude formula for zero-frequency conductivity\cite{PhysRevLett.2000.Mayou, JMathPhys.1997.Roche, PhysRevLett.1997.Schulz-Baldes}
 \begin{equation}
  \label{equ:drude}
  \sigma \simeq  e^2 \varrho(E_\mathrm{F}) c t_\mathrm{sc}^{2\beta -1} \;.
  \end{equation}
The quantity $e$ denotes the elementary charge of an electron, $\varrho(E_\mathrm{F})$ the density of states at the Fermi level $E_\mathrm{F}$, $c$ a constant, and $t_\mathrm{sc}$ a characteristic time beyond which propagation becomes diffusive due to scattering.\cite{PhysRevLett.2000.Mayou, RevMathPhys.1998.SchulzBaldes} Hence, $\beta=0$ corresponds to the absence of diffusion, $\beta=1/2$ to classical diffusion, and $\beta=1$ to ballistic spreading. For quasiperiodic structures one often observes anomalous diffusion characterized by $0 < \beta < 1 $.\cite{PhysRevB.1992.Passaro, JPhysFrance.1989.Sire, PhysRevB.2000.Yuan}

Within the range of anomalous diffusion, the sub-diffusive regime for $\beta < \tfrac{1}{2}$ is most interesting because in this case the conductivity $\sigma$ in Eq.\ \eqref{equ:drude} decreases with increasing scattering time $t_\mathrm{sc}$. This can be interpreted as a system for which the particles are trapped due to quantum interference phenomena and conductivity is mainly caused by non-elastic collisions, where the collision rate increases with decreasing scattering time $t_\mathrm{sc}$.\cite{PhysRevLett.1997.Schulz-Baldes} For $\beta > \tfrac{1}{2}$ the reverse behavior can be observed, i.e., the conductivity decreases with increasing collision rate. We already pointed out that various experiments for real quasicrystals revealed an electronic transport in agreement to the conductivity in the subdiffusive regime ($\beta < \tfrac{1}{2}$).\cite{PhysRevLett.1993.Mayou, AdvPhys.1992.Poon, JMathPhys.1997.Roche}

\begin{figure}
  \centering
  \includegraphics[width=\columnwidth]{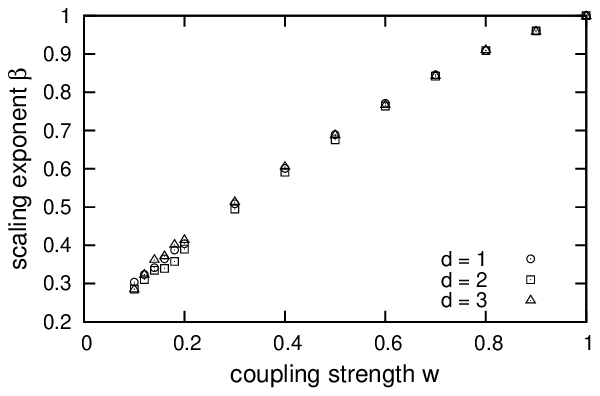}
  \includegraphics[width=\columnwidth]{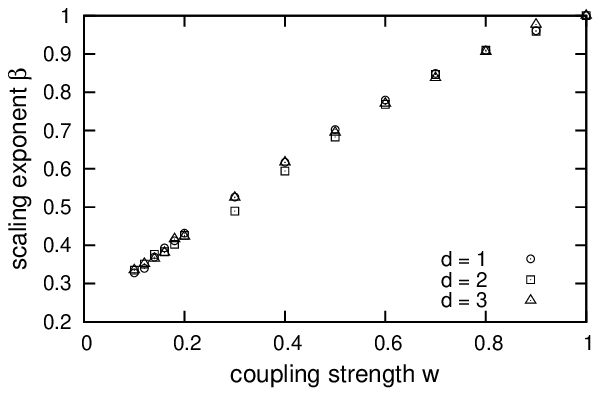}
  \caption{Scaling exponents $\beta$ for the golden-mean (top) and the silver-mean (bottom) systems in one, two, and three dimensions for $s=1$. All results are averaged over different initial positions of the wave packet.}
  \label{fig:betanD}
\end{figure}

The scaling behavior of the mean square displacement $d(t)$ is visualized for the silver-mean chain $\mathcal{C}^\mathrm{Ag}$ in Fig.\ \ref{fig:msd-octonacci} for $s = 1$ and different coupling strengths $w$. The results are averaged over 300 initial positions of the wave packet around the center of the chain. We observe a scaling behavior according to Eq.\ \eqref{equ:msd-scaling} over several orders of magnitude till the width becomes constant due to finite size effects. The corresponding scaling exponents $\beta$ for the average mean square displacement $d(t)$ are shown in Fig.\ \ref{fig:betanD} for different metallic-mean chains. Thereby, the scaling exponents $\beta$ steadily increase with the coupling parameter $w$ and the parameter $b$ of the inflation rule.

In Fig.\ \ref{fig:msd-octonacci} the mean square displacement $d(t)$ is also visualized for the two- and three-dimensional silver-mean labyrinth tiling, where we averaged in two dimensions over 50 and in three dimensions over 20 initial positions of the wave packet due to the limitations of computing resources. The results are qualitatively very similar to that of the quasiperiodic chains and the scaling exponents in Fig.\ \ref{fig:betanD} even show that the one-dimensional scaling exponents $\beta^\mathrm{1d}$ are nearly identical to the results for the two- and three-dimensional labyrinth tiling.

\section{RG Approaches for Golden- and Silver-Mean Chains}
\label{sec:RG-theory}

\begin{figure}[t!]
 \centering
 \footnotesize
 \includegraphics[width=\columnwidth]{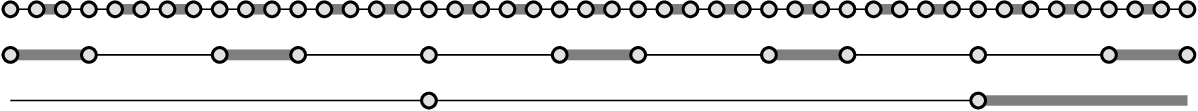}\vspace{0.5cm}
 \includegraphics[width=\columnwidth]{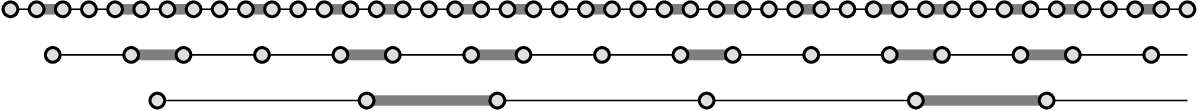}
  \caption{Atomic RG (top) and molecular RG (bottom) for the Fibonacci chain $\mathcal{C}^\mathrm{Au}$: in the atomic RG all atomic sites survive an RG step, and in the molecular RG all molecular sites are replaced by a new site. The thick (thin) lines denote the strong (weak) bonds.}
 \label{fig:rg-fib}
\end{figure}

The structure and the dynamical properties of the metallic-mean systems can be described by an RG approach, which is a mathematical apparatus for the systematical study of the changes of a physical system viewed at different length scales. The general idea is that a given system can be transformed (renormalized) in an RG step in such a way that we obtain a new system with a reduced number of degrees of freedom (i.e.\ sites and bonds). We make use of the fact that the metallic-mean chains contain for $w = 0$ only isolated sites (atoms) and biatomic clusters (molecules) coupled by a strong bond as visualized in Fig.\ \ref{fig:rg-fib}. This yields three highly degenerate energy levels: $E = 0$ for the atomic sites and $E = \pm s$ for the bonding and antibonding states of the molecules.\cite{PhysRevLett.1986.Niu} For non-zero parameters $w$ one finds a coupling between these isolated clusters, where the dominant contribution occurs between sites of the same type.\cite{PhysRevB.1990.Niu}  For $w \ll s$ also the probability density of the wave functions is concentrated either on atomic or on molecular sites depending on the energy $E$. For the Fibonacci chain $\mathcal{C}^\mathrm{Au}$ Niu and Nori\cite{PhysRevLett.1986.Niu} distinguished two possible RG approaches:

\emph{Atomic RG} --- The renormalized grid after one RG step consists of the atomic sites of the original chain as shown in Fig.\ \ref{fig:rg-fib}. These new sites are then connected by new strong and weak bonds. In particular, atoms separated by one molecule in the original chain get connected by a new strong bond and atoms separated by two molecules in the original chain become connected by a new weak bond (cf.\ also Fig.\ \ref{fig:rg-fib-atom2}).

\emph{Molecular RG} --- In one RG step all molecular sites of the original chain are replaced by new atomic sites (cf.\ Fig.\ \ref{fig:rg-fib}). Again the atoms in the renormalized grid can be connected by new bonds, where the sites of neighboring molecules are connected by a new strong bond and sites of molecules separated by an atomic site become connected by a new weak bond (cf.\ also Fig.\ \ref{fig:rg-fib-mol2}).

Both RG schemes yield a new Fibonacci chain, which is scaled in length and energy. Basically, the substitution of a cluster of bonds by a new bond during an RG step corresponds to the block diagonalization of the Hamiltonian of the original chain. This yields an effective Hamiltonian for the new chain containing the bond strengths between the remaining sites.\cite{PhysRevB.1990.Niu} The derivation of the scaling factors $z$ for the new bond strengths as well as the length scalings $c$ under the application of one RG step is described in detail in the Appendix \ref{subsec:atomic-rg} for the atomic RG and Appendix \ref{subsec:molecular-rg} for the molecular RG approach.

\begin{figure}[b!]
 \centering
 \footnotesize
 \includegraphics[width=\columnwidth]{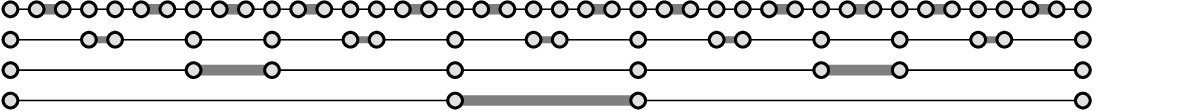}\vspace{0.5cm}
 \includegraphics[width=\columnwidth]{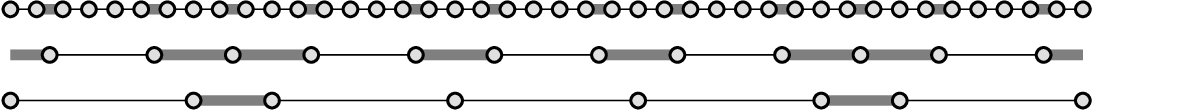}
 \caption{Atomic RG (top) and molecular RG (bottom) for the silver-mean chain $\mathcal{C}^\mathrm{Ag}$: in the atomic RG all atoms survive and for the molecular RG all molecular sites are replaced by a new site, which results in a new silver-mean chain only after two steps.}
 \label{fig:rg-oct}
\end{figure}

For the silver-mean chain $\mathcal{C}^\mathrm{Ag}$ we can apply the same approach. In the atomic RG the renormalized grid again consists of all atomic sites of the original chain (cf.\ Fig.\ \ref{fig:rg-oct}). By appropriately assigning weak and strong bonds we obtain a new silver-mean chain, which is scaled in length and energy. However, the molecular RG approach for the silver-mean chain is more complicated and requires a two-step process to yield a new silver-mean chain as shown in Fig.\ \ref{fig:rg-oct}. The new silver-mean chain contains then all atomic sites which are located between two molecular sites of the original chain. Details are presented in Appendix \ref{subsec:octonacci-rg-atom} for the atomic RG and Appendix \ref{subsec:octonacci-rg-mol} for the molecular RG approach.

\section{RG Approach for Wave-packet Dynamics in Metallic-Mean Chains}
\label{sec:RG-theory-msd}

The scaling exponent $\beta$ of the mean square displacement $d(t)$ for the golden-mean and the silver-mean chain in the regime of strong quasiperiodic modulations ($w \ll s$) can be determined by an approach proposed by Abe and Hiramoto\cite{PhysRevA.1987.Abe,JPhysJap.1988.Hiramoto} and further refined by Pi\'echon.\cite{PhysRevLett.1996.Piechon}
It is based on the idea that the time evolution of an initially localized wave packet can be considered as the realizations of successive steps of the above described RG. In particular, this leads to the following two transformations:

\emph{Length scaling} --- In one RG step the width $d(t)$ of the wave packet scales with the grid spacing $c$, i.e., $d(t) \to d^\prime(t) / c$.

\emph{Time scaling} --- The time evolution of a wave packet is governed by the hierarchic structure of the tiling, which leads to the occurrence of hierarchic  resonances on different time scales (cf.\ Fig.\ \ref{fig:rg-oct-atom}).\cite{JPhysJap.1988.Hiramoto, PhysRevB.2009.Thiem, PhysRevB.1994.Wilkinson} Since for $w \ll s$ the energy scales of these resonances are very different, no interference of electrons on different time scales is considered. Further, the scaling factor for the time $t$ under one RG step equals the inverse of the scaling factor $z = 1/t$ of the energy. The second factor is obtained by computing the strengths of the new bonds of the different clusters in the RG approach by applying Brillouin-Wigner (BW) perturbation theory, as discussed in the appendix.

Therefor, from the scaling behavior $d(t) \simeq d_0 t^\beta$ of the mean square displacement according to Eq.\ \eqref{equ:msd-scaling}, one can derive the relation\cite{PhysRevA.1987.Abe,JPhysJap.1988.Hiramoto}
    \begin{equation}
        \label{equ:rg-beta}
        \beta \simeq \frac{\ln c}{\ln z} \;.
    \end{equation}

Abe and Hiramoto as well as Pi\'echon presented only results for the Fibonacci chain for the two cases of the atomic RG and the molecular RG.\cite{PhysRevA.1987.Abe, JPhysJap.1988.Hiramoto, PhysRevLett.1996.Piechon} In this section we also derive an approximation of the scaling exponent $\beta$ for the silver-mean chain and show for the associated labyrinth tilings that the corresponding scaling exponents approach the one-dimensional results in the regime of strong quasiperiodic fluctuations. The analytical expressions are also compared with numerical results, which show a very good agreement.

\mathversion{bold}
\subsection{Fibonacci chain $\mathcal{C}^\mathrm{Au}$}
\mathversion{normal}

Using the results by Niu and Nori\cite{PhysRevLett.1986.Niu, PhysRevB.1990.Niu} (cf.\ Appendix \ref{subsec:atomic-rg}), the energy scaling for the weak and the strong clusters in the atomic RG is $\bar{z}_\mathrm{Au} = {w^2}/{s^2}$. With the length scaling $c_\mathrm{Au}^\mathrm{atom} = \tau_\mathrm{Au}^{-3}$ they obtained for $\beta$ the expression \cite{JPhysJap.1988.Hiramoto}
\begin{equation}
 \label{equ:rg-atom-golden-mean}
 \beta_\mathrm{Au}^\mathrm{atom} \simeq \frac{3 \ln \tau_\mathrm{Au}}{2 \ln \frac{s}{w}}\;.
\end{equation}

For the molecular RG the energy scaling for the weak and the strong clusters is $z_\mathrm{Au} = {w}/{2s}$  and the length scaling $c_\mathrm{Au}^\mathrm{mol} = \tau_\mathrm{Au}^{-2}$. This yields the scaling exponent \cite{JPhysJap.1988.Hiramoto}
\begin{equation}
 \label{equ:rg-mol-golden-mean}
 \beta_\mathrm{Au}^\mathrm{mol} \simeq \frac{2 \ln \tau_\mathrm{Au}}{\ln \frac{s}{w} + \ln 2}\;.
\end{equation}
While the results by Abe and Hiramoto are based on a quantitative argument, rigorous mathematical considerations by Damanik\cite{PhilMag.2006.Damanik, JAmMath.2007.Damanik} showed that for a periodic chain with a quasiperiodic potential (diagonal model) modulated according to the Fibonacci sequence a very similar relation can be found for $w \ll s$. With the constants $c_1$ and $c_2$ this result is given by $c_1 / \ln \tfrac{s}{w} < \beta < c_2 / \ln \tfrac{s}{w}$.

\begin{figure}[t!]
  \centering
  \includegraphics[width=\columnwidth]{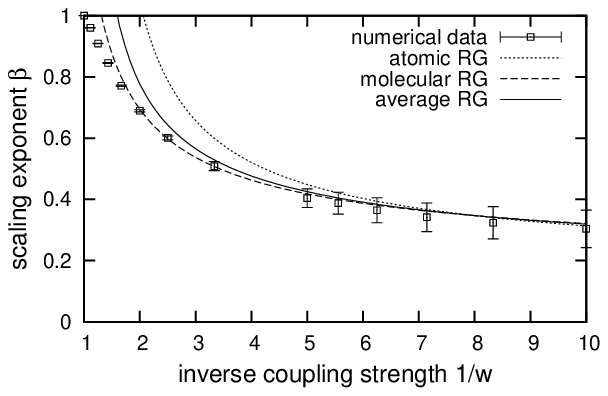}\hspace{-0.2cm}
  \includegraphics[width=\columnwidth]{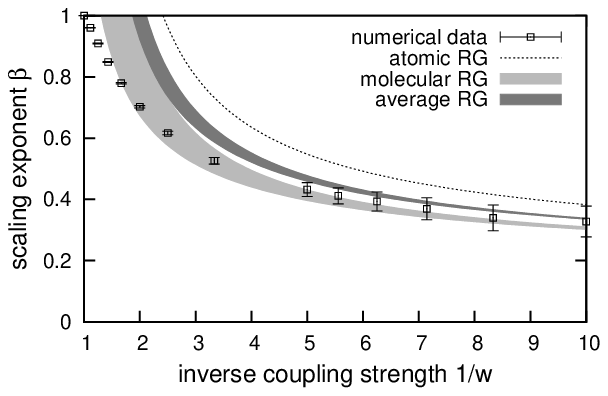}
  \caption{Scaling exponent $\beta$ of the mean square displacement $d(t)$ averaged over different initial positions in comparison to the analytical results of the atomic and molecular RG approach for the golden-mean (top) and the silver-mean systems (bottom) in one, two, and three dimensions for $s=1$.}
  \label{fig:rg-plot1D}
\end{figure}

In Fig.\ \ref{fig:rg-plot1D} both scaling exponents, $\beta_\mathrm{Au}^\mathrm{atom}$ and $\beta_\mathrm{Au}^\mathrm{mol}$, are compared with the numerical results of the scaling exponent $\beta$ of the mean square displacement $d(t)$. While for small coupling parameters $w$ the results are very close for both RG approaches, the molecular RG theory yields significantly better results for the scaling exponent $\beta$ for larger values of $w$. This can be understood by the fact that the dynamics of an initially localized wave packet are actually governed by the atomic \emph{and} the molecular RG approach.\cite{PhysRevLett.1996.Piechon,JPhysJap.1988.Hiramoto} The reason is that independently of the original type (atomic or molecular site) of the initial position $l_0$ of the wave packet after one RG step the new initial site $l_0^\prime$ can be either an atomic or a molecular site. Thus, the scaling exponent $\beta$ of the mean square displacement $d(t)$ shown in Fig.\ \ref{fig:rg-plot1D} strongly depends on the percentages $p_\mathrm{atom}$ and $p_\mathrm{mol}$ of atoms and molecules in the golden-mean chain. As the percentage $p_\mathrm{mol}$ is significantly higher, it is not surprising that the exponent $\beta_\mathrm{Au}^\mathrm{mol}$ shows a better agreement with the numerical data for the exponent $\beta$. One can also determine a weighted average of both scaling exponents, that is,
\begin{subequations}
\label{equ:average-beta-fib}
\begin{align}
   \beta_\mathrm{Au}
    &\stackrel{\hphantom{N_a \to \infty}}{=} p_\mathrm{atom} \beta_\mathrm{Au}^\mathrm{atom} + p_\mathrm{mol} \beta_\mathrm{Au}^\mathrm{mol} \\
    & \stackrel{N_a \to \infty}{=} \frac{  \beta_\mathrm{Au}^\mathrm{atom} + 2 \#_s(\mathcal{C}_a^\mathrm{Au}) \left(\beta_\mathrm{Au}^\mathrm{mol} - \beta_\mathrm{Au}^\mathrm{atom}\right) }{N_a} \\
    &\stackrel{\substack{\hphantom{N_a \to \infty}\\{\eqref{equ:tau}}}}{=} \frac{\tau_\mathrm{Au}-1}{\tau_\mathrm{Au}+1} \beta_\mathrm{Au}^\mathrm{atom} + \frac{2}{\tau_\mathrm{Au}+1} \beta_\mathrm{Au}^\mathrm{mol} \;.
\end{align}
\end{subequations}
A comparison of this expression with the numerical results in Fig.\ \ref{fig:rg-plot1D} shows a very good agreement for $w \ll s$, but for larger $w$ the agreement is not as good as for $\beta_\mathrm{Au}^\mathrm{mol}$. In the second case we have to keep in mind that we apply perturbation theory and the results are only valid for $w \ll s$. As the RG approach overestimates the results for larger values of $w$, usually the lower of the two exponents, $\beta_\mathrm{Au}^\mathrm{atom}$ and $\beta_\mathrm{Au}^\mathrm{mol}$, yields the better results for $w \ge 0.3 s$.

It is also possible to determine the scaling exponents, $\beta_\mathrm{Au}^\mathrm{atom}$ and $\beta_\mathrm{Au}^\mathrm{mol}$, numerically. This is achieved by studying wave packets which are initially localized at a position $l_0$ that survives all atomic/molecular RG steps for an approximant of finite size. Using this method Hiramoto and Abe found a good correspondence between the theoretical and numerical results of the atomic RG for $w < s/4$.\cite{JPhysJap.1988.Hiramoto}

\mathversion{bold}
\subsection{Silver-mean chain $\mathcal{C}^\mathrm{Ag}$}
\mathversion{normal}

The derivations for the atomic RG approach in Appendix~\ref{subsec:octonacci-rg-atom} lead for both types of clusters to an energy scaling of $\bar{z}_\mathrm{Ag} = w/s$. With the corresponding length scaling $c_\mathrm{Ag}^\mathrm{atom} = \tau_\mathrm{Ag}^{-1}$ we obtain the scaling exponent
\begin{equation}
 \label{equ:rg-atom-silver-mean}
 \beta_\mathrm{Ag}^\mathrm{atom} \simeq \frac{\ln \tau_\mathrm{Ag}}{\ln \frac{s}{w}}\;.
\end{equation}

For the molecular RG the energy scaling for the weak and strong cluster is different. In Appendix \ref{subsec:octonacci-rg-mol} we obtain for the strong cluster $z_\mathrm{Ag}^\mathrm{sc} = {w^2}/{s^2}$ and for the weak cluster $z_\mathrm{Ag}^\mathrm{wc} < {w^3}/{s^3}$ for $w \ll s$. On average we expect to obtain a superposition of the spreading caused by both clusters. Using Eq.\ \eqref{equ:tau}, which relates the number of strong and weak bonds/clusters in the silver-mean chain, we can derive an average energy scaling of
\begin{subequations}
 \label{equ:zeff-oct}
\begin{align}
 z_\mathrm{Ag} &= \frac{1}{N_a} \left ( \#_s(\mathcal{C}_a) z_\mathrm{Ag}^\mathrm{sc} + \#_w(\mathcal{C}_a) z_\mathrm{Ag}^\mathrm{wc} \right)\\
 &       = \frac{1}{1 + \tau_\mathrm{Ag}} \frac{w^2}{s^2} +  \frac{\tau_\mathrm{Ag}}{1+\tau_\mathrm{Ag}} z_\mathrm{Ag}^\mathrm{wc} \;.
\end{align}
\end{subequations}
With the length scaling of $c_\mathrm{Ag}^\mathrm{mol} = \tau_\mathrm{Ag}^{-2}$ we obtain for the scaling exponent $\beta_\mathrm{Ag}^\mathrm{mol}$ the expression
\begin{equation}
 \label{equ:rg-mol-silver-mean}
 \beta_\mathrm{Ag}^\mathrm{mol} \simeq \frac{2 \ln \tau_\mathrm{Ag}}{\ln  \left(1+\tau_\mathrm{Ag}\right) -  \ln \left( \tfrac{w^2}{s^2} + \tau_\mathrm{Ag} z_\mathrm{Ag}^\mathrm{wc} \right) }\;.
\end{equation}

\begin{figure}[b!]
  \centering
  \includegraphics[width=\columnwidth]{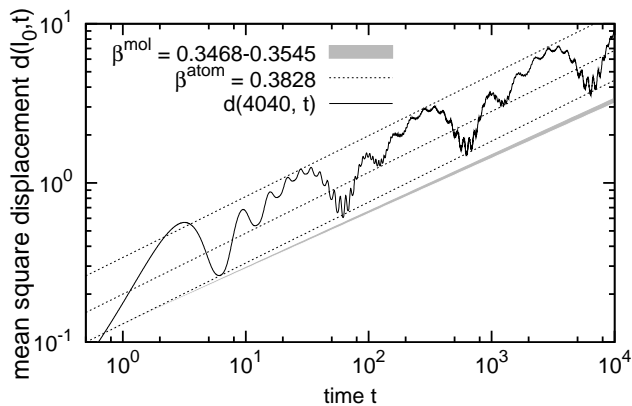}
  \caption{Mean-square displacement $d(l_0=4040,t)$ for the silver-mean chain $\mathcal{C}_{11}^\mathrm{Ag}$ for $w=0.1$ and $s=1$. The initial site $l_0$ of the wave packet remains an atomic site for the first five atomic RG steps. The slope is well described by $\beta^\mathrm{atom}$.}
  \label{fig:rg-oct-atom}
\end{figure}

In Fig.\ \ref{fig:rg-plot1D} the analytical expressions for the atomic and the molecular RG are compared with the numerical results. For the molecular RG we always plot a range of possible scaling exponents $\beta_\mathrm{Ag}^\mathrm{mol}$ by choosing $z_\mathrm{Ag}^\mathrm{wc} \in [0,w^3/{s^3}]$. The results show that the scaling exponent $\beta_\mathrm{Ag}^\mathrm{mol}$ slightly underestimates the numerical values for $w \ll s$. In contrast, the analytical values $\beta_\mathrm{Ag}^\mathrm{atom}$ of the atomic RG are significantly higher than the numerical results.
As already discussed for the golden-mean chain, the  exponent $\beta$ depends on the initial position $l_0$ of the wave packet. For instance, Fig.\ \ref{fig:rg-oct-atom} shows that an atomic initial position $l_0$ which remains an atomic site under the RG transformation is well described by the atomic RG. Hence, on average we expect to obtain a scaling exponent $\beta$ between the results predicted by the atomic and the molecular RG. Using the percentages of atoms and molecules in the silver-mean chain according to Eq.\ \eqref{equ:tau}, we obtain for the average scaling exponent
\begin{equation}
 \label{equ:rgAv-silver-mean}
 \beta_\mathrm{Ag} \simeq \frac{\tau_\mathrm{Ag}-1}{\tau_\mathrm{Ag}+1}  \beta_\mathrm{Ag}^\mathrm{atom}  + \frac{2}{\tau_\mathrm{Ag}+1}  \beta_\mathrm{Ag}^\mathrm{mol}
\end{equation}
in correspondence with Eq.\ \eqref{equ:average-beta-fib}.
This expression is visualized in Fig.\ \ref{fig:rg-plot1D} as well. The results show that the analytical values correspond to the numerical values within the error bars for $w \ll s$. But again, the lower exponent $\beta_\mathrm{Ag}^\mathrm{mol}$ shows a better agreement for larger $w$. Further, by comparing the results for the golden-mean and the silver-mean chain we find that the exponent $\beta_\mathrm{Ag}$ is slightly larger than $\beta_\mathrm{Au}$ in agreement with the numerical results in Fig.\ \ref{fig:betanD}.

Further, the repeating patterns of $d(t)$ in Fig.\ \ref{fig:rg-oct-atom} reflect the hierarchic resonances of the wave packet dynamics on different time scales. We can also determine the scaling factor $z$ directly from the distance of successive patterns. In the case of the atomic RG for the silver-mean chain for $w=0.1$ and $s=1$ the patterns are repeated in Fig.\ \ref{fig:rg-oct-atom} after a factor of 10 on the time scale, which yields the scaling factor $\bar{z}_\mathrm{Ag} = 1/10$ for the energy and, thus, for the bond strengths. This result agrees perfectly with the theoretical result $\bar{z} = w/s$.

\begin{figure*}[t!]
 \centering
 \includegraphics[width=0.33\textwidth]{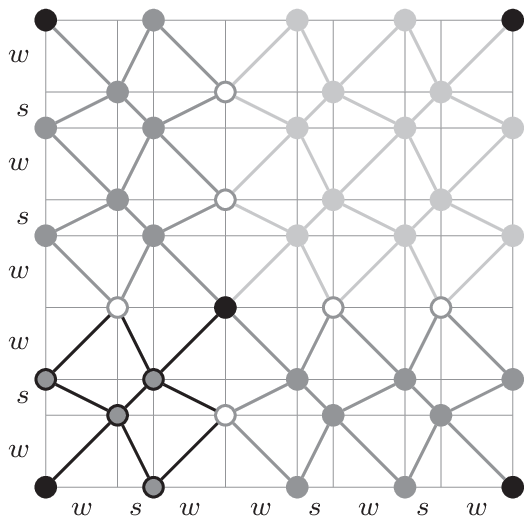}\hfill
 \includegraphics[width=0.33\textwidth]{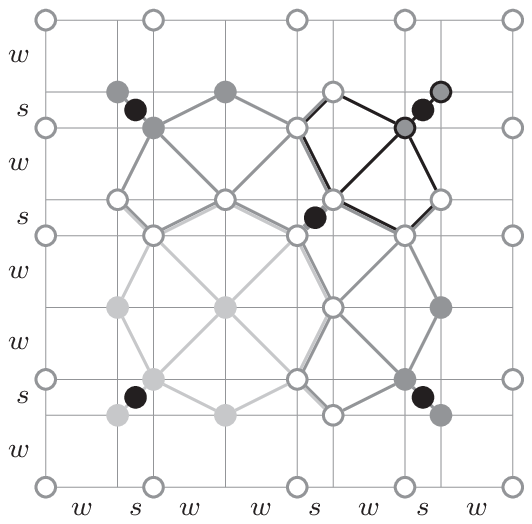}\hfill
 \includegraphics[width=0.33\textwidth]{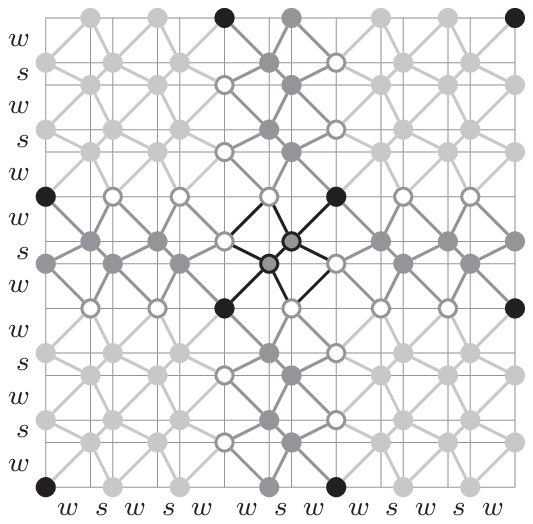}
 \caption{Strong (black lines), medium (dark gray lines), and weak (light gray lines) clusters for the atomic RG (left) and the molecular RG (center) for the golden-mean labyrinth tiling $\mathcal{L}^\mathrm{Ag}$. Sites of the renormalized grid are shown in black and sites belonging to more than one cluster in white. In the right panel we show a part of the golden-mean labyrinth tiling $\mathcal{L}^\mathrm{Au}$ which after one atomic RG step results in the strong cluster in the left panel.}
  \label{fig:rg-labyrinth}
\end{figure*}

\section{RG Theory for Higher Dimensions}
\label{sec:rg-labyrinth}

Like the one-dimensional systems, the labyrinth tilings possess a hierarchical structure. In Fig.\ \ref{fig:rg-labyrinth} we show the corresponding atomic and the molecular RG approach for the golden-mean labyrinth tiling $\mathcal{L}^\mathrm{Au}$. In this tiling we can identify three different clusters (strong, medium, and weak) in correspondence to the three different bond types. In general, the clusters of the RG in $d$ dimensions are given by the product of the corresponding one-dimensional clusters due to the product structure of the labyrinth tiling (cf.\ Figs.\ \ref{fig:rg-labyrinth-atom-strong} to \ref{fig:rg-labyrinth-mol-weak}). Hence, the number of different clusters depends on the number of bond types, which in $d$ dimensions is given by $d+1$.

The derivations of an analytical expression for the scaling of the length and bonds strengths for the labyrinth tiling follow the same way as before. As an example they are described for the golden-mean labyrinth tiling $\mathcal{L}^\mathrm{Au}$ in the Appendices \ref{subsec:labyrinth-fib-atom} and \ref{subsec:labyrinth-fib-mol}. In two dimensions we obtain a grid scaling of $c_\mathrm{Au}^\mathrm{atom,2d} = \tau_\mathrm{Au}^{-6}$ for the atomic RG and $c_\mathrm{Au}^\mathrm{mol,2d} = \tau_\mathrm{Au}^{-4}$ for the molecular RG. With the BW perturbation theory we determine the new bond strengths, where we obtain a qualitatively different result for the atomic and the molecular RG.

For the atomic RG the scaling factor for the bond strength is $\bar{z}_\mathrm{Au}^\mathrm{2d} = \left(\bar{z}_\mathrm{Au} \right)^2 = {w^4}/{s^4}$. Plugging this into Eq.\ \eqref{equ:rg-beta} we obtain the same analytical expression as in one dimension, i.e., Eq.\ \eqref{equ:rg-atom-golden-mean}. This result is not surprising because the scaling factors for the grid spacing and the bond strengths are the squares of the one-dimensional results, which cancel each other according to Eq.\ \eqref{equ:rg-beta}. Since the dominant coupling between the edge sites of a $d$-dimensional cluster originates from the bonds along its diagonal and the couplings along this diagonal are the products of the coupling strengths of $d$ one-dimensional clusters (cf.\ Figs.\ \ref{fig:rg-labyrinth-atom-strong} to \ref{fig:rg-labyrinth-atom-weak}), for the atomic RG in $d$ dimensions the scaling factor for the bond strengths and, hence, the energies is $\bar{z}^d$.

\begin{figure}[b!]
  \centering
  \includegraphics[width=\columnwidth]{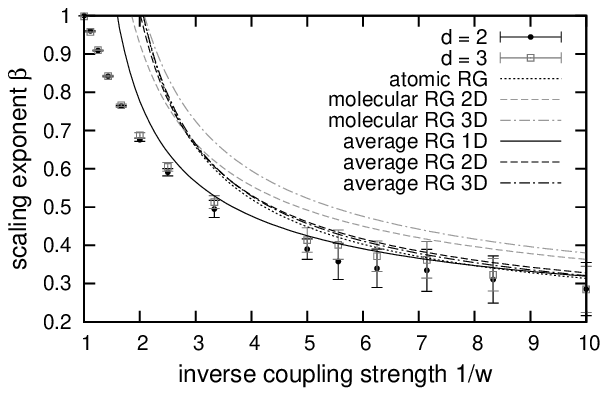}
  \includegraphics[width=\columnwidth]{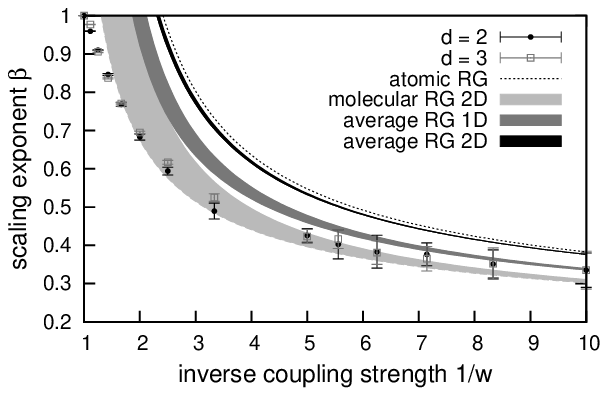}
  \caption{Scaling exponent $\beta$ of the mean square displacement $d(t)$ averaged over different initial positions in comparison to the analytical results of the atomic and molecular RG approach for the golden-mean labyrinth tiling $\mathcal{L}^\mathrm{Au}$ (top) and the silver-mean labyrinth tiling $\mathcal{L}^\mathrm{Ag}$ (bottom) for $s=1$.}
  \label{fig:rg-plot2D}
\end{figure}

Although the structure of the clusters is very similar for the molecular RG, the result is somewhat different than the one-dimensional exponent in Eq.\ \eqref{equ:rg-mol-golden-mean}. The derivations in Appendix \ref{subsec:labyrinth-fib-mol} yield a scaling of $z_\mathrm{Au}^\mathrm{2d} = 2 \left(z_\mathrm{Au} \right)^2 = {w^2}/{2s^2}$ with an additional factor of $2$ originating from the normalization condition. Hence, we obtain
\begin{equation}
 \label{equ:rg-mol-golden-mean-2D}
 \beta_\mathrm{Au}^{\mathrm{mol},2\mathrm{d}} \simeq \frac{2 \ln \tau_\mathrm{Au}}{\ln \frac{s}{w} + \frac{1}{2} \ln 2}\;.
\end{equation}
For very small coupling parameters $w \to 0$ the first term of the denominator dominates and this expression approaches the one-dimensional result in Eq.\ \eqref{equ:rg-mol-golden-mean}. We plot this function in Fig.\ \ref{fig:rg-plot2D} and find that it is greater than the one-dimensional results of the molecular RG for all considered values of $w$. For the average scaling behavior we have to consider the contributions of both RG approaches according to Eqs.\ \eqref{equ:rg-atom-golden-mean} and \eqref{equ:rg-mol-golden-mean-2D}. By calculating the percentages of atoms and molecules, we obtain the average scaling exponent
\begin{subequations}
   \label{equ:average-beta-fib-2d}
\begin{align}
   \beta_\mathrm{Au}^\mathrm{2d}
    &\stackrel{\hphantom{\eqref{equ:tau}}}{=} p_\mathrm{atom}^\mathrm{2d} \beta_\mathrm{Au}^{\mathrm{atom},2\mathrm{d}} + p_\mathrm{mol}^\mathrm{2d} \beta_\mathrm{Au}^{\mathrm{mol},2\mathrm{d}} \\
   & \stackrel{\eqref{equ:tau}}{=}  \beta_\mathrm{Au}^{\mathrm{atom},2\mathrm{d}} + \frac{2  \left( \beta_\mathrm{Au}^{\mathrm{mol},2\mathrm{d}} - \beta_\mathrm{Au}^{\mathrm{atom},2\mathrm{d}}  \right)  }{\left(\tau_\mathrm{Au}+1\right)^2}\;.
\end{align}
\end{subequations}

The result for the three-dimensional labyrinth tiling $\mathcal{L}^\mathrm{3d, Au}$ follows in the same way. We already pointed out that the atomic RG approach yields the same exponent in every dimension (cf.\ Eq.\ \eqref{equ:rg-atom-golden-mean}). Due to the normalization condition we obtain in the molecular RG an additional factor of $4$, which yields the scaling factor $z_\mathrm{Au}^\mathrm{3d} = 4 \left(z_\mathrm{Au} \right)^3 = {w^3}/{2s^3}$. Hence, with $c_\mathrm{Au}^\mathrm{mol,3d} = \tau_\mathrm{Au}^{-6}$ the analytical expression for the molecular RG is $\beta_\mathrm{Au}^{\mathrm{mol},3\mathrm{d}} \simeq \left( 2 \ln \tau_\mathrm{Au} \right) / \left( \ln ({s}/{w}) + \frac{1}{3} \ln 2 \right)$. This yields the average scaling exponent
\begin{align}
\label{equ:average-beta-fib-3d}
   \beta_\mathrm{Au}^\mathrm{3d}
    &\stackrel{\eqref{equ:tau}}{=} \beta_\mathrm{Au}^{\mathrm{atom},3\mathrm{d}} + \frac{2 \left( \beta_\mathrm{Au}^{\mathrm{mol},3\mathrm{d}} - \beta_\mathrm{Au}^{\mathrm{atom},3\mathrm{d}} \right) }{\left(\tau_\mathrm{Au}+1\right)^3} \;.
\end{align}
In Fig.\ \ref{fig:rg-plot2D} we compare the numerical and the analytical results for the average scaling exponent $\beta_\mathrm{Au}$ for the golden-mean labyrinth tilings. Within the error bounds we find a good correspondence in the regime of strong quasiperiodic modulation. Further, although the analytical expressions for the scaling exponent $\beta_\mathrm{Au}$ in different dimensions are quite different, we see that they approach each other for $w \ll s$. This is also in good agreement with the numerical data in Fig.\ \ref{fig:betanD}, which show no significant differences in one, two, and three dimensions.

\begin{figure}[b!]
 \footnotesize
 \centering
 \includegraphics[width=0.49\columnwidth]{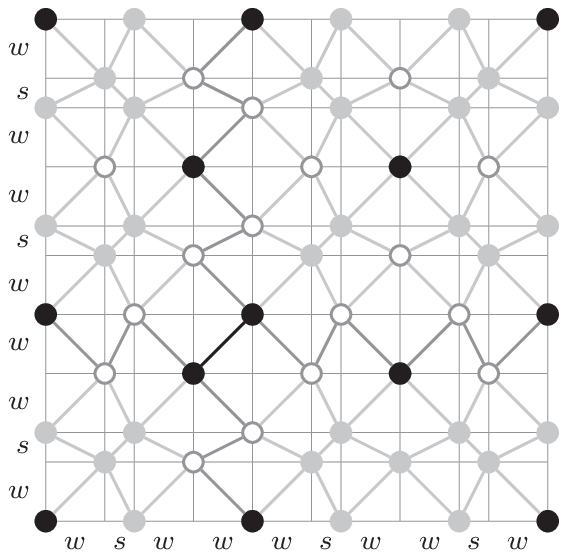}
 \includegraphics[width=0.49\columnwidth]{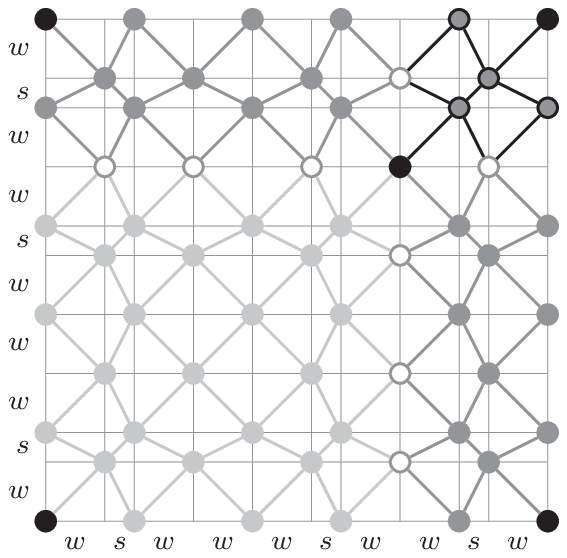}
 \caption{Strong (black lines), medium (dark gray lines), and weak (light gray lines) clusters for the atomic RG (left) and the molecular RG (right) for the silver-mean labyrinth tiling $\mathcal{L}^\mathrm{Ag}$. Sites of the renormalized grid are shown in black.}
  \label{fig:rg-labyrinth-silver}
\end{figure}

Also for the silver-mean labyrinth tiling $\mathcal{L}^\mathrm{Ag}$ we can apply the atomic and molecular RG. Like for the golden-mean labyrinth tiling we find three types of clusters in two dimensions, which are visualized in Fig.\ \ref{fig:rg-labyrinth-silver}. Again, we have to distinguish between the atomic and the molecular RG approach. For the atomic RG the scaling factors of the length as well as the bond strengths are the squares of the one-dimensional results, which yields the same analytical expression for $\beta$ (cf.\ Eq.\ \eqref{equ:rg-atom-silver-mean}) as in one dimension.

For the molecular RG we have to carefully adapt the RG approach because the scaling factors of the energies are not identical for the different types of clusters. Hence, we explicitly have to determine the new average scaling factor $z_\mathrm{Ag}$ (cf.\ Eq.\ \eqref{equ:zeff-oct}). The scaling factors for the clusters are given as the product of the one-dimensional results, which yields for the strong, medium, and weak clusters of the two-dimensional labyrinth tiling the factors $z_\mathrm{Ag}^\mathrm{sc,2d} = w^4/s^4$, $z_\mathrm{Ag}^\mathrm{mc,2d} = z_\mathrm{Ag}^\mathrm{wc} w^2/s^2 $, and $z_\mathrm{Ag}^\mathrm{wc,2d} = \left(z_\mathrm{Ag}^\mathrm{wc}\right)^2$, respectively. By considering the percentage of each of these clusters in the labyrinth tiling, we obtain the average scaling factor
\begin{subequations}
 \label{equ:zeff-oct-2d}
\begin{align}
 z_\mathrm{Ag}
  &= \frac{1}{N_a^2} \left( \#_s(\mathcal{C}_a) \right)^2 z_\mathrm{Ag}^\mathrm{sc,2d} + \frac{2}{N_a^2} \#_s(\mathcal{C}_a) \#_w(\mathcal{C}_a) z_\mathrm{Ag}^\mathrm{mc,2d} \nonumber\\
  &\quad\; + \frac{1}{N_a^2} \left( \#_w(\mathcal{C}_a) \right)^2 z_\mathrm{Ag}^\mathrm{wc,2d}  \\
  &=         \frac{1}{\left(1 + \tau_\mathrm{Ag}\right)^2} \left( \frac{w^2}{s^2} +  \tau_\mathrm{Ag} z_\mathrm{Ag}^\mathrm{wc} \right)^2 \;.
\end{align}
\end{subequations}
According to Eq.\ \eqref{equ:rg-beta} this yields the same scaling exponent as for the one-dimensional system (cf.\ Eq.\ \eqref{equ:rg-mol-silver-mean}).
The average scaling exponent $\beta_\mathrm{Ag}$ follows then from Eq.\ \eqref{equ:average-beta-fib-2d} by replacing the quantities for the golden-mean labyrinth tiling with that of the silver-mean labyrinth tiling. Since we only know the upper bound of the scaling factor of the weak cluster in one dimension (i.e.\ $z_\mathrm{Ag}^\mathrm{wc} \in [0,w^3/s^3]$), the analytical expressions again describe a range of possible exponents. A comparison with the numerical results in Fig.\ \ref{fig:rg-plot2D} shows that the final scaling exponent is very close to that of the atomic RG and, hence, overestimates the actual results.
Rather, the molecular RG works better than the average.
By performing the same calculations also for three dimensions, we obtain that the scaling exponent $\beta_\mathrm{Ag}^\mathrm{3d}$ is even closer to the scaling exponent of the atomic RG. Hence, we would expect that also the numerical results for $\beta$ approach the analytical expression for $\beta_\mathrm{Ag}^\mathrm{atom}$ in Eq.\ \eqref{equ:rg-atom-silver-mean} with increasing dimensionality. However, we do not observe such a behavior.  A possible explanation for these discrepancies could be that determining an average scaling factor for the bond strength in the molecular RG according to Eq.\ \eqref{equ:zeff-oct-2d} from the energy scalings of the different cluster types does not adequately display the dynamical behavior. At least, the numerical and analytical results approach each other with increasing quasiperiodic modulation, and we find a correspondence for $w=0.1$ within the error bounds.

\section{Conclusion}
\label{sec:conclusion}

We have studied the quantum diffusion in quasiperiodic systems in one, two, and three dimensions by investigating the time evolution of wave packets.
By numerical calculations of the scaling behavior of the mean square displacement $d(t)$ of a wave packet we have observed the occurrence of anomalous transport for all coupling strengths $w \in (0,s)$ and that the scaling exponents $\beta$ are more or less independent of the dimension.

Further, we have extended an RG method, originally proposed by Abe and Hiramoto for the golden-mean chain to obtain an analytical expression also for the scaling exponent $\beta$ of the silver-mean chain. With the same method we have been able to show that the scaling exponents $\beta^{d\mathrm{d}}$ of the labyrinth tilings approach the one-dimensional scaling exponents $\beta^{1\mathrm{d}}$ for the golden-mean and the silver-mean system in the regime of strong quasiperiodic modulation ($w \ll s$). The analytical results are also in good agreement to our numerical results.

Although properties like the structure of the energy spectrum strongly depend on the dimensionality, these derivations also showed that the product structure is an essential reason for the rather similar scaling exponents $\beta$ in different dimensions. Hence, further research of the transport properties is needed especially with a focus on non-separable tilings. Nevertheless, we have been able to draw a connection between the structure of the quasiperiodic systems and their transport properties with the RG approach. We found that for strong quasiperiodic modulations the wave-packet dynamics are governed by the hierarchical structure of the quasiperiodic systems leading to the occurrence of hierarchic resonances on different time scales.

Additionally, for $w \ll s$ the numerical and analytical results for the exponent $\beta$ show the occurrence of sub\-diffusive wave-packet dynamics ($\beta < 1/2$). According to the generalized Drude equation this leads to a decrease of the conductivity with increasing structural order as it is observed in real quasicrystals.\cite{JMathPhys.1997.Roche, PhysRevLett.1993.Mayou} Hence, labyrinth tilings can be useful models to study the characteristics of higher-dimensional quasiperiodic systems efficiently by numerical methods and also analytically.

\appendix
\section{Renormalization of lengths and energies of quasiperiodic systems}
\label{app:renormalization}

This section comprises the results for the scaling of lengths and energies in the RG approach introduced by Niu and Nori.\cite{PhysRevLett.1986.Niu, PhysRevB.1990.Niu} Although the scaling factors have been derived for the Fibonacci chain before,\cite{PhysRevA.1987.Abe, JPhysJap.1988.Hiramoto} we will briefly sketch the calculation for the atomic RG and mention the results for the molecular RG because we make use of them also for the octonacco chain. Further, it allows us to compare the results for different dimensions.

\mathversion{bold}
\subsection{Fibonacci chain $\mathcal{C}^\mathrm{Au}$ --- Atomic RG}
\label{subsec:atomic-rg}
\mathversion{normal}

\begin{figure}[b!]
 \footnotesize
  \includegraphics[scale=1]{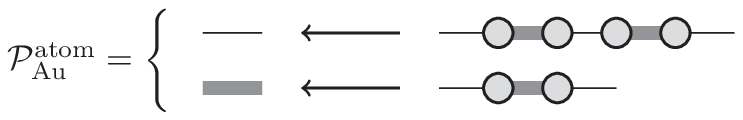}\\\vspace{0.3cm}
  \includegraphics[scale=1]{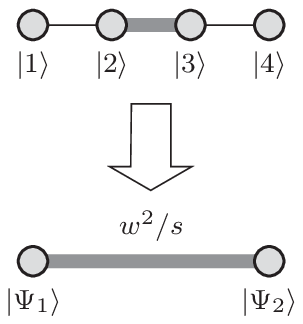}\hspace{0.2cm}
  \includegraphics[scale=1]{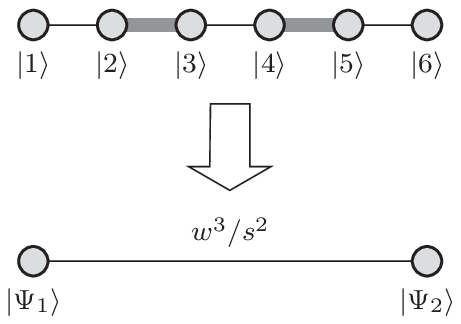}\\
 \caption{Substitution rule $\mathcal{P}_\mathrm{Au}^\mathrm{atom}$ of the atomic RG for the Fibonacci chain $\mathcal{C}^\mathrm{Au}$ shown in Fig.\ \ref{fig:rg-fib} (top) with the energy scaling for the strong cluster (left) and the weak cluster (right).}
 \label{fig:rg-fib-atom2}
\end{figure}

In the atomic RG approach for the Fibonacci chain the renormalized grid after one RG step consists of the atomic sites of the original chain (cf.\ Fig.\ \ref{fig:rg-fib}), which leads to a scaling of the bonds according to the substitution rule in Fig.\ \ref{fig:rg-fib-atom2}. The scaling of the grid spacing is given by the substitution matrix\cite{ModPhys.1993.Baake} of the corresponding RG expansion $(\mathcal{P}_\mathrm{Au}^\mathrm{atom})^{-1}$, which relates the number of symbols $w$ and $s$ in successive RG steps:
\begin{equation}
 \mathbf{R}\left( (\mathcal{P}_\mathrm{Au}^\mathrm{atom} )^{-1}\right) =
  \begin{pmatrix}
  \#_w(w_w) & \#_s(w_w) \\
  \#_w(w_s) & \#_s(w_s)
 \end{pmatrix} =
 \begin{pmatrix}
  3 & 2 \\
  2 & 1
 \end{pmatrix}.
\end{equation}
The Perron-Frobenius eigenvalue $\lambda_\mathrm{Fr}$ of this matrix corresponds to the inverse of the scaling factor $c_\mathrm{Au}^\mathrm{atom}$ of the grid spacing, i.e.,
\begin{align}
 \det( \mathbf{R} - \lambda \mathbf{I} ) &= \lambda^2 - 4 \lambda - 1 \stackrel{!}{=} 0  \nonumber \\
 \Longrightarrow c_\mathrm{Au}^\mathrm{atom} &= \frac{1}{\lambda_\mathrm{Fr}} = \frac{1}{2 + \sqrt{5}} = \tau_\mathrm{Au}^{-3} \;.
\end{align}

The scaling of the new bond strengths is calculated by Brillouin-Wigner (BW) perturbation theory for each of the two substitution clusters in Fig.\ \ref{fig:rg-fib-atom2}. For the metallic-mean systems the coupling parameter $w$ is treated as a perturbation. Hence, the Hamiltonian $\mathbf{H} = \mathbf{H}_0 + \mathbf{H}_1$ is decomposed into a Hamiltonian $\mathbf{H}_0(w=0)$ of the unperturbed system and a perturbation $\mathbf{H}_1 = \mathbf{H} - \mathbf{H}_0(w=0)$. The approach is described in detail by Niu and Nori.\cite{PhysRevB.1990.Niu}

\subsubsection{Energy scaling for strong cluster}

The Hamiltonian of the strong cluster in Fig.\ \ref{fig:rg-fib-atom2} is
\begin{equation}
\mathbf{H} = \mathbf{H}_0 + \mathbf{H}_1
 =  \begin{pmatrix}
  0 & 0 & 0 & 0 \\
  0 & 0 & s & 0 \\
  0 & s & 0 & 0 \\
  0 & 0 & 0 & 0
 \end{pmatrix} +
 \begin{pmatrix}
  0 & w & 0 & 0 \\
  w & 0 & 0 & 0 \\
  0 & 0 & 0 & w \\
  0 & 0 & w & 0
 \end{pmatrix} \;.
\end{equation}
Solving the Schr\"odinger equation for the unperturbed Hamiltonian $\mathbf{H}_0$ yields the wave functions $\ket{\Psi_1} = \ket{1}$, $\ket{\Psi_2} = \ket{4}$, and $\ket{\Psi_3^\pm} = \tfrac{1}{\sqrt{2}}\left(\ket{2} \pm \ket{3} \right)$ with their corresponding energy values $E_{1/2} = 0$, and $E_3^\pm = \pm s$.

The strength $s^\prime$ of the new bond is given by the matrix element $\bra{\Psi_1}\mathbf{H}\ket{\Psi_2}$, where only the leading non-zero term of the perturbation expansion is relevant. The BW perturbation theory yields no contribution in the first order because $\bra{\Psi_1}\mathbf{H}_1\ket{\Psi_2} = 0$. In the second order we obtain a new coupling strength of
\begin{subequations}
 \label{equ:rg-fib-atom-strong}
\begin{align}
 &\bra{\Psi_1}\mathbf{H}_1 \mathbf{P} \frac{1}{E-\mathbf{H}_0}\mathbf{P}\mathbf{H}_1\ket{\Psi_2} \\
 &= \bra{\Psi_1}\mathbf{H}_1 \ket{\Psi_3^+} \frac{1}{E_1-E_3^+} \bra{\Psi_3^+}\mathbf{H}_1\ket{\Psi_2} + \nonumber \\
 & \quad\; \bra{\Psi_1}\mathbf{H}_1 \ket{\Psi_3^-} \frac{1}{E_1-E_3^-} \bra{\Psi_3^-}\mathbf{H}_1\ket{\Psi_2} \\
 & = 2 \frac{w}{\sqrt{2}} \frac{-w}{\sqrt{2}s} = - \frac{w^2}{s} \\
 & \Longrightarrow s^\prime = \frac{w^2}{s} = \frac{w^2}{s^2} s  \;.
\end{align}
\end{subequations}
Here, $\mathbf{P}$ denotes the projection operator out of the subspace for a given $g$-times degenerate eigenstate $\Psi_{i}$ with the energy $E_{i}$ of $\mathbf{H}_0$, i.e., for the strong cluster it is given by $\mathbf{P} = \ket{\Psi_3^+}\bra{\Psi_3^+} + \ket{\Psi_3^-}\bra{\Psi_3^-}$.

\subsubsection{Energy scaling for weak cluster}

Analogously this approach is applied to the weak cluster.
The Hamiltonian $\mathbf{H}_0$ yields the atomic wave functions $\ket{\Psi_1} = \ket{1}$, $\ket{\Psi_2} = \ket{6}$ with the energy $E_{1/2} = 0$ as well as the molecular wave functions $\ket{\Psi_3^\pm} = \frac{1}{\sqrt{2}}\left(\ket{2} \pm \ket{3} \right)$ and $\ket{\Psi_4^\pm} = \frac{1}{\sqrt{2}}\left(\ket{4} \pm \ket{5} \right)$ with the energies $E_{3/4}^\pm = \pm s$.

The strength $w^\prime$ of the new bond is again given by the matrix element $\bra{\Psi_1}\mathbf{H}\ket{\Psi_2}$. We obtain no contribution in the first order expansion due to $\bra{\Psi_1}\mathbf{H}_1\ket{\Psi_2} = 0$ and with the projection operator $\mathbf{P} = \ket{\Psi_3^+}\bra{\Psi_3^+} + \ket{\Psi_3^-}\bra{\Psi_3^-} + \ket{\Psi_4^+}\bra{\Psi_4^+} + \ket{\Psi_4^-}\bra{\Psi_4^-}$ also the second order term vanishes, $\bra{\Psi_1} \mathbf{H}_1 \mathbf{P} \frac{1}{E-\mathbf{H}_0}\mathbf{P}\mathbf{H}_1\ket{\Psi_2} = 0$. In the third order we have four non-zero terms:
{\allowdisplaybreaks
\begin{subequations}
\begin{align}
 &\bra{\Psi_1}\mathbf{H}_1 \mathbf{P} \frac{1}{E-\mathbf{H}_0}\mathbf{P}\mathbf{H}_1\mathbf{P} \frac{1}{E-\mathbf{H}_0}\mathbf{P}\mathbf{H}_1\ket{\Psi_2} \\
 &= \sum \bra{\Psi_1}\mathbf{H}_1 \ket{\Psi_3^\pm} \frac{1}{E_1-E_3^\pm} \bra{\Psi_3^\pm}\mathbf{H}_1\ket{\Psi_4^\pm} \nonumber \\
 & \quad\; \frac{1}{E_1-E_4^\pm} \bra{\Psi_4^\pm}\mathbf{H}_1\ket{\Psi_2} \\
 &= 4\frac{w}{\sqrt{2}} \frac{-w}{2s} \frac{-w}{\sqrt{2}s} = \frac{w^3}{s^2} \\
 & \Longrightarrow  w^\prime = \frac{w^3}{s^2} = \frac{w^2}{s^2} w \;.
\end{align}
\end{subequations}}

Hence, for the atomic RG of the Fibonacci chain the scaling of the bond strength of both clusters and, thus, the scaling of the energy is given by $\bar{z}_\mathrm{Au} = {w^2}/{s^2}$.

\mathversion{bold}
\subsection{Fibonacci chain $\mathcal{C}^\mathrm{Au}$ --- Molecular RG}
\mathversion{normal}
\label{subsec:molecular-rg}

\begin{figure}[t!]
 \footnotesize
 \centering
  \includegraphics[scale = 1]{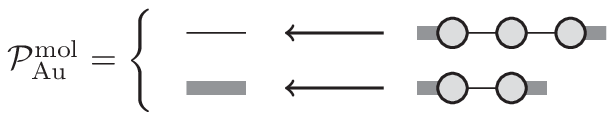}\\\vspace{0.26cm}
  \includegraphics[scale = 1]{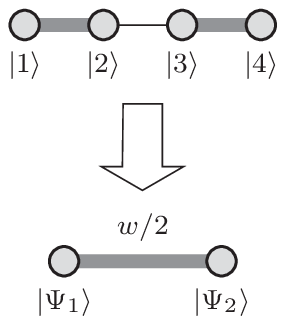}\hspace{0.2cm}
  \includegraphics[scale = 1]{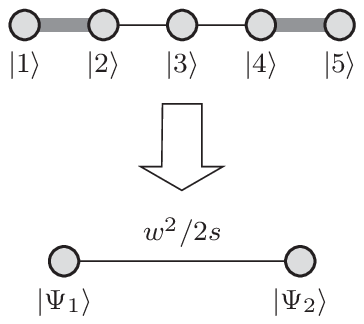}
 \caption{Substitution rule $\mathcal{P}_\mathrm{Au}^\mathrm{mol}$ of the molecular RG for the Fibonacci chain $\mathcal{C}^\mathrm{Au}$ in Fig.\ \ref{fig:rg-fib} (top) with the energy scaling for the strong cluster (left) and the weak cluster (right).}
 \label{fig:rg-fib-mol2}
\end{figure}

In the molecular RG of the Fibonacci chain all molecular sites of the original chain are replaced by new atomic sites as shown in Fig.\ \ref{fig:rg-fib}, which leads to a scaling of the bonds according to the substitution rule $\mathcal{P}_\mathrm{Au}^\mathrm{mol}$ in Fig.\ \ref{fig:rg-fib-mol2}. The scaling factor $c_\mathrm{Au}^\mathrm{mol}$ of the grid spacing is again determined by the substitution matrix
\begin{equation}
 \mathbf{R}\left( \left(\mathcal{P}_\mathrm{Au}^\mathrm{mol}\right)^{-1}\right) =
 \begin{pmatrix}
  2 & 1 \\
  1 & 1
 \end{pmatrix}\quad,
\end{equation}
which results in
\begin{align}
  c_\mathrm{Au}^\mathrm{mol} &= \frac{1}{\lambda_\mathrm{Fr}} = \frac{2} { 3 + \sqrt{5}} = \tau_\mathrm{Au}^{-2}  \;.
\end{align}

The scaling factors for the new bonds of the two substitution clusters in Figs.\ \ref{fig:rg-fib-mol2} follow by BW perturbation theory. The derivations are described in detail by Niu und Nori \cite{PhysRevB.1990.Niu} and yield
\begin{subequations}
\begin{align}
 s^\prime &= \frac{w}{2} = \frac{w}{2s} s\\
 w^\prime &= \frac{w^2}{2s} = \frac{w}{2s} w  \;.
\end{align}
\end{subequations}
This corresponds to an energy scaling $z_\mathrm{Au} = {w}/{2s}$.

\mathversion{bold}
\subsection{Octonacci chain $\mathcal{C}^\mathrm{Ag}$ --- Atomic RG}
\mathversion{normal}
\label{subsec:octonacci-rg-atom}

\begin{figure}[b!]
 \centering
  \includegraphics[scale = 1]{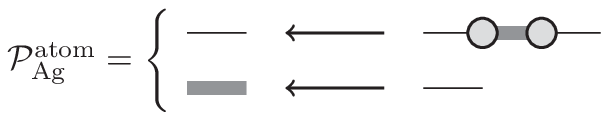}\\\vspace{0.2cm}
  \includegraphics[scale = 0.9]{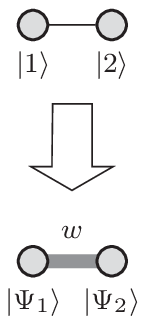}\hspace{1cm}
  \includegraphics[scale = 0.9]{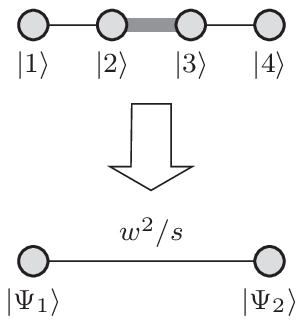}
 \caption{Substitution rule $\mathcal{P}_\mathrm{Ag}^\mathrm{atom}$ of the atomic RG for the silver-mean chain $\mathcal{C}^\mathrm{Ag}$ in Fig.\ \ref{fig:rg-oct} (top) with the energy scaling for the strong cluster (left) and the weak cluster (right).}
 \label{fig:rg-oct-atom2}
\end{figure}

Like in the atomic RG approach for the Fibonacci chain, here only atomic sites survive during an RG step as shown in Fig.\ \ref{fig:rg-oct}. The corresponding substitution rule and the scaling of the clusters are shown in Fig.\ \ref{fig:rg-oct-atom2}. The substitution matrix is now given by
\begin{equation}
 \mathbf{R}\left( \left(\mathcal{P}_\mathrm{Ag}^\mathrm{atom} \right)^{-1} \right) =
 \begin{pmatrix}
  2 & 1 \\
  1 & 0
 \end{pmatrix}\quad,
\end{equation}
and thus the grid spacing scales with
\begin{align}
 c_\mathrm{Ag}^\mathrm{atom} &=  \frac{1} {\lambda_\mathrm{Fr}} = \frac{1}{1 + \sqrt{2}} = \tau_\mathrm{Ag}^{-1} \;.
\end{align}

The bond strength for the strong cluster is very simple, and for the weak cluster (cf.\ Fig.\ \ref{fig:rg-oct-atom2}) we already derived its strength in Eq.\ \eqref{equ:rg-fib-atom-strong} while studying the strong cluster in the atomic RG of the Fibonacci chain (cf.\ Fig.\ \ref{fig:rg-fib-atom2}), i.e.,
\begin{subequations}
\begin{align}
 s^\prime &= w = \frac{w}{s} s \\
 w^\prime &= \frac{w^2}{s} = \frac{w}{s} w  \;.
\end{align}
\end{subequations}
Hence, we obtain an energy scaling $\bar{z}_\mathrm{Ag} = {w}/{s}$.

\mathversion{bold}
\subsection{Octonacci chain $\mathcal{C}^\mathrm{Ag}$ --- Molecular RG}
\mathversion{normal}
\label{subsec:octonacci-rg-mol}

\begin{figure}[b!]
 \centering
  \includegraphics[scale = 1]{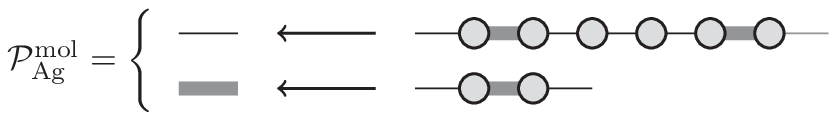}\\\vspace{0.2cm}
  \includegraphics[scale = 0.9]{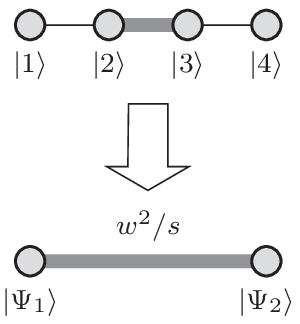}\hspace{0.2cm}
  \includegraphics[scale = 0.9]{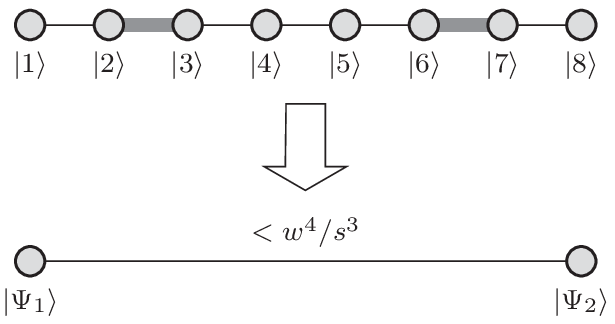}
 \caption{Substitution rule $\mathcal{P}_\mathrm{Ag}^\mathrm{mol}$ of the molecular RG for the silver-mean chain  $\mathcal{C}^\mathrm{Ag}$ in Fig.\ \ref{fig:rg-oct} (top) with the energy scaling for the strong cluster (left) and the weak cluster (right).}
 \label{fig:rg-oct-mol2}
\end{figure}

Applying the molecular RG to the silver-mean chain $\mathcal{C}^\mathrm{Ag}$ leads to some challenges because two RG steps are needed to obtain a new silver-mean chain (cf.\ Fig.\ \ref{fig:rg-oct}). The substitution matrix of the corresponding RG expansion (cf.\ Fig.\ \ref{fig:rg-oct-mol2}) is given by
\begin{equation}
 \mathbf{R}\left( \left( \mathcal{P}_\mathrm{Ag}^\mathrm{mol} \right)^{-1} \right) =
 \begin{pmatrix}
  5 & 2 \\
  2 & 1
 \end{pmatrix}
\quad,
\end{equation}
and we obtain a scaling of the grid spacing of
\begin{align}
 c_\mathrm{Ag}^\mathrm{mol} &=  \frac{1}{\lambda_\mathrm{Fr}} = \frac{1}{3 + 2\sqrt{2}} = \tau_\mathrm{Ag}^{-2} \;.
\end{align}
The scaling of the bond strengths is again calculated using the BW perturbation theory for each of the two substitution clusters in Fig.\ \ref{fig:rg-oct-mol2}.

\subsubsection{Energy scaling for strong cluster}

Here, we have the same substitution rule as for the strong cluster in the atomic RG theory for the Fibonacci chain (cf.\ Fig.\ \ref{fig:rg-fib-atom2}), that is, we get
\begin{equation}
s^\prime = \frac{w^2}{s} = \frac{w^2}{s^2} s
\end{equation}
This results in an energy scaling of $z_\mathrm{Ag}^\mathrm{sc} = {w^2}/{s^2}$ for the strong cluster.

\subsubsection{Energy scaling for weak cluster}

The computation of the strength of the new weak bond turns out to be rather complicated. The reason is that we cannot apply the method used so far because the edge sites of the cluster and some of the inner sites belong to the same subspace. Hence, we have to resolve this degeneracy first to compute the new Hamiltonian matrix elements. However, the new set of wave functions shows already a coupling of the edge sites of the cluster so that we do not obtain information about the energy scaling of the bonds.

However, we can obtain an upper bound replacing the central weak bond of the weak cluster by a strong bond. For this modified cluster we can apply the RG approach. In particular, we obtain for the Hamiltonian $\mathbf{H}_0$ of the modified cluster the atomic eigenstates $\ket{\Psi_1} = \ket{1}$, $\ket{\Psi_2} = \ket{8}$ with $E_{1/2} = 0$ and the molecular eigenstates $\ket{\Psi_3^\pm} = \frac{1}{\sqrt{2}}\left(\ket{2} \pm \ket{3} \right)$, $\ket{\Psi_4^\pm} = \frac{1}{\sqrt{2}}\left(\ket{4} \pm \ket{5} \right)$, $\ket{\Psi_5^\pm} = \frac{1}{\sqrt{2}}\left(\ket{6} \pm \ket{7} \right)$ with $E_{3-5}^\pm = \pm s$.

The leading non-zero contribution occurs only in the fourth order of the perturbation expansion:
{\allowdisplaybreaks
\begin{subequations}
\begin{align}
 &\bra{\Psi_1}\mathbf{H}_1  ( \mathbf{P}  \frac{1}{E-\mathbf{H}_0}\mathbf{P}\mathbf{H}_1 )^3 \ket{\Psi_2} \\
 &=\sum \bra{\Psi_1}\mathbf{H}_1 \ket{\Psi_3^\pm} \frac{1}{E_1 - E_3^\pm} \bra{\Psi_3^\pm}\mathbf{H}_1 \ket{\Psi_4^\pm} \frac{1}{E_1-E_4^\pm} \nonumber \\
 &\quad \;  \bra{\Psi_4^\pm}\mathbf{H}_1\ket{\Psi_5^\pm} \frac{1}{E_1 - E_5^\pm} \bra{\Psi_5^\pm}\mathbf{H}_1 \ket{\Psi_2}\\
 &= 8 \frac{w}{\sqrt{2}} \left(\frac{-w}{2s} \right)^2 \frac{-w}{\sqrt{2}s}  = - \frac{w^4}{s^3} \\
 &\Longrightarrow  w^\prime = \frac{w^3}{s^3} w \;.
\end{align}
\end{subequations}}
This new cluster possesses a larger coupling between the edge states than the original cluster shown in Fig.\ \ref{fig:rg-oct-mol2}. Hence, the corresponding scaling factor provides an upper bound for the scaling factor of the weak cluster with $z_\mathrm{Ag}^\mathrm{wc} < {w^3}/{s^3}$.

\mathversion{bold}
\section{Renormalization of Lengths and Energies for the Golden-Mean Labyrinth $\mathcal{L}^\mathrm{Au}$}

\subsection{Atomic RG}
\mathversion{normal}
\label{subsec:labyrinth-fib-atom}

\begin{figure}[b!]
  \centering
  \includegraphics[width=0.75\columnwidth]{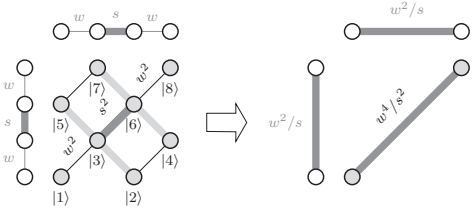}
 \caption{Substitution rule for the strong cluster in the atomic RG approach of the golden-mean labyrinth tiling $\mathcal{L}^\mathrm{Au}$.}
 \label{fig:rg-labyrinth-atom-strong}
\end{figure}

For the labyrinth tiling we use the same technique as for the quasiperiodic chains. For the atomic RG the renormalized labyrinth tiling contains only the atomic sites of the original tiling as shown in Fig.\ \ref{fig:rg-labyrinth} for the golden-mean labyrinth tiling $\mathcal{L}^\mathrm{Au}$. In two dimensions we obtain three types of clusters. The corresponding substitution rules for the strong, medium, and weak cluster are shown in Figs.\ \ref{fig:rg-labyrinth-atom-strong}, \ref{fig:rg-labyrinth-atom-medium}, and \ref{fig:rg-labyrinth-atom-weak}. The scaling of the grid spacing $c$ can be easily obtained from the scaling $N_a^\prime \to c N_a$ of the one-dimensional system size. Thus, we obtain in $d$ dimensions
\begin{equation}
    \label{equ:rg-length-scaling-nD}
    V_a^\prime = \frac{1}{2^{d-1}} \left(N_a^\prime\right)^d  = \frac{1}{2^{d-1}} c^d N_a^d = c^d V_a \;.
\end{equation}
Hence, for the atomic RG in two dimensions the scaling of the grid spacing is
\begin{equation}
 c_\mathrm{Au}^\mathrm{atom,2d} = \tau_\mathrm{Au}^{-6} \;.
\end{equation}

\subsubsection{Energy scaling for strong cluster}

We assign each vertex $(l,m)$ of the two-dimensional clusters a unique index $r = \lceil{\tfrac{l}{2}}\rceil+ \lceil{(m-1)\tfrac{f}{2}}\rceil$, where $f$ is the length of the underlying one-dimensional cluster. This corresponds to the row-wise assignment of successive indices beginning from the lower left corner to the upper right corner. With this mapping we can write the wave functions as a vector and the Hamiltonian of the strong cluster of the RG is given by
\begin{equation}
\mathbf{H} =
 \begin{pmatrix}
    0   & 0     & w^2   & 0     & 0     & 0     & 0     & 0 \\
    0   & 0     & sw    & w^2   & 0     & 0     & 0     & 0 \\
    w^2 & sw    & 0     & 0     & sw    & s^2   & 0     & 0 \\
    0   & w^2   & 0     & 0     & 0     & sw    & 0     & 0 \\
    0   & 0     & sw    & 0     & 0     & 0     & w^2   & 0 \\
    0   & 0     & s^2   & sw    & 0     & 0     & sw    & w^2\\
    0   & 0     & 0     & 0     & w^2   & sw    & 0     & 0 \\
    0   & 0     & 0     & 0     & 0     & w^2   & 0     & 0
 \end{pmatrix} \quad.
\end{equation}
The corresponding Hamiltonian $\mathbf{H}_0$ yields the normalized wave functions $\ket{\Psi_1} = \ket{1}$, $\ket{\Psi_2} = \ket{2}$, $\ket{\Psi_3} = \ket{4}$, $\ket{\Psi_4} = \ket{5}$, $\ket{\Psi_5} = \ket{7}$, $\ket{\Psi_6} = \ket{8}$,  and $\ket{\Psi_7^\pm} = \frac{1}{\sqrt{2}}\left(\ket{3} \pm \ket{6} \right)$ with the energy values $E_{1-6} = 0$ and $E_7^\pm = \pm s^2$.

The strength of the new bond is given by the matrix element $\bra{\Psi_1}\mathbf{H}\ket{\Psi_6}$. In analogy to the one-dimensional case, the first order contribution is $\bra{\Psi_1}\mathbf{H}_1\ket{\Psi_6} = 0$, and the second order contribution yields with $\mathbf{P} = \ket{\Psi_7^+}\bra{\Psi_7^+} + \ket{\Psi_7^-}\bra{\Psi_7^-}$ the non-zero term
{\allowdisplaybreaks
\begin{subequations}
\begin{align}
 \bra{\Psi_1}&\mathbf{H}_1 P \frac{1}{E-\mathbf{H}_0}\mathbf{P}\mathbf{H}_1\ket{\Psi_6}  \\
 &= \bra{\Psi_1}\mathbf{H}_1 \ket{\Psi_7^+} \frac{1}{E_1-E_7^+} \bra{\Psi_7^+}\mathbf{H}_1\ket{\Psi_6} + \nonumber \\
 & \quad\; \bra{\Psi_1}\mathbf{H}_1 \ket{\Psi_7^-} \frac{1}{E_1-E_7^-} \bra{\Psi_7^-}\mathbf{H}_1\ket{\Psi_6} \\
 &= 2 \frac{w^2}{\sqrt{2}} \frac{w^2}{-\sqrt{2}s^2} = - \frac{w^4}{s^2} \\
 & \Longrightarrow \left(s^\prime\right)^2 = \frac{w^4}{s^2} = \frac{w^4}{s^4} s^2  \;.
\end{align}
\end{subequations}}

\subsubsection{Energy scaling for medium cluster}

\begin{figure}[t!]
  \centering
  \includegraphics[width=\columnwidth]{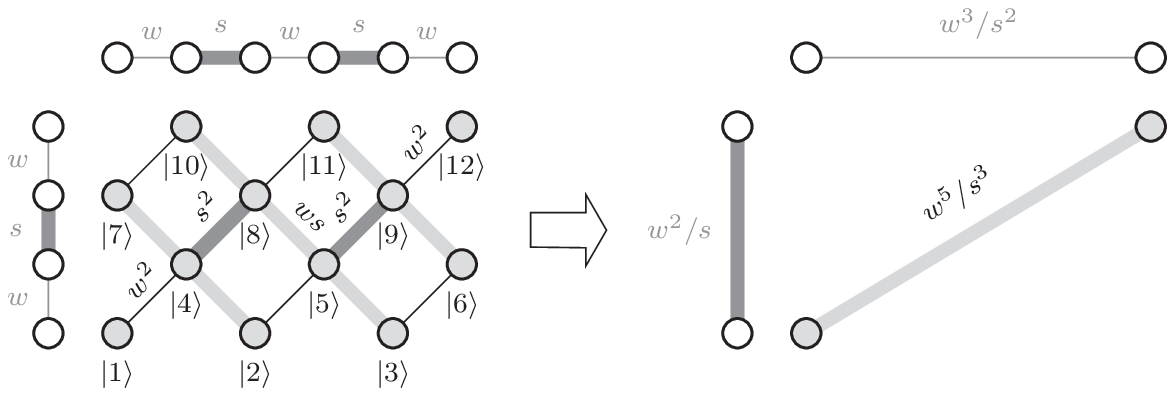}
 \caption{Substitution rule for the medium cluster in the atomic RG approach of the golden-mean labyrinth tiling $\mathcal{L}^\mathrm{Au}$.}
 \label{fig:rg-labyrinth-atom-medium}
\end{figure}

For the cluster of medium strength we use the same approach. The Hamiltonian again contains all couplings as shown in Fig.\ \ref{fig:rg-labyrinth-atom-medium} and the eigenstates of $\mathbf{H}_0$ are $\ket{\Psi_1} = \ket{1}$, $\ket{\Psi_2} = \ket{2}$, $\ket{\Psi_3} = \ket{3}$, $\ket{\Psi_4} = \ket{6}$, $\ket{\Psi_5} = \ket{7}$, $\ket{\Psi_6} = \ket{10}$, $\ket{\Psi_7} = \ket{11}$, $\ket{\Psi_8} = \ket{12}$ with $E_{1-8} = 0$ and $\ket{\Psi_{9}^\pm} = \frac{1}{\sqrt{2}}\left(\ket{4} \pm \ket{8} \right)$, $\ket{\Psi_{10}^\pm} = \frac{1}{\sqrt{2}}\left(\ket{5} \pm \ket{9} \right)$ with $E_{9/10}^\pm = \pm s^2$.

The new bond strength of the medium cluster is given by the matrix element $\bra{\Psi_1}\mathbf{H}\ket{\Psi_{8}}$. Thus, there is no contribution in the first and the second order expansion.
In the third order we have four contributing terms:
\begin{subequations}
\begin{align}
 &\bra{\Psi_1}\mathbf{H}_1 \mathbf{P} \frac{1}{E-\mathbf{H}_0}\mathbf{P}\mathbf{H}_1\mathbf{P} \frac{1}{E-\mathbf{H}_0}\mathbf{P}\mathbf{H}_1\ket{\Psi_{8}} \\
 &= \sum \bra{\Psi_1}\mathbf{H}_1 \ket{\Psi_{9}^\pm} \frac{1}{E_1-E_{9}^\pm} \bra{\Psi_{9}^\pm}\mathbf{H}_1\ket{\Psi_{10}^\pm} \nonumber\\
 & \quad\; \frac{1}{E_1-E_{10}^\pm} \bra{\Psi_{10}^\pm}\mathbf{H}_1\ket{\Psi_{8}} \\
 &= 4 \frac{w^2}{\sqrt{2}} \frac{ws}{-2s^2} \frac{w^2}{-\sqrt{2}s^2}
 = \frac{w^5}{s^3} \\
 &\Longrightarrow  w^\prime s^\prime = \frac{w^5}{s^3} = \frac{w^4}{s^4} ws
\end{align}
\end{subequations}

\subsubsection{Energy scaling for weak cluster}

\begin{figure}[t!]
  \centering
  \includegraphics[width=\columnwidth]{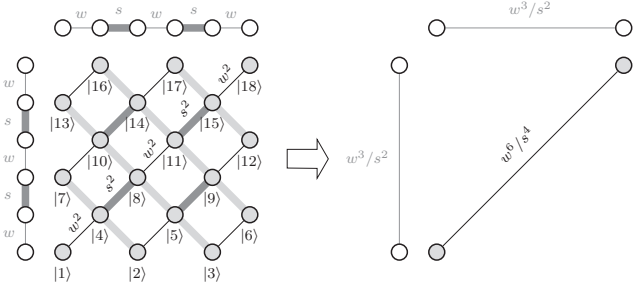}
  \caption{Substitution rule for the weak cluster in the atomic RG approach of the golden-mean labyrinth tiling $\mathcal{L}^\mathrm{Au}$.}
 \label{fig:rg-labyrinth-atom-weak}
\end{figure}

For the weak cluster the Hamiltonian again contains all coupling strengths as shown in Fig.\ \ref{fig:rg-labyrinth-atom-weak}. The Hamiltonian $\mathbf{H}_0$ yields the atomic wave functions $\ket{\Psi_1} = \ket{1}$, $\ket{\Psi_2} = \ket{2}$, $\ket{\Psi_3} = \ket{3}$, $\ket{\Psi_4} = \ket{6}$, $\ket{\Psi_5} = \ket{7}$, $\ket{\Psi_6} = \ket{12}$, $\ket{\Psi_7} = \ket{13}$, $\ket{\Psi_8} = \ket{16}$, $\ket{\Psi_9} = \ket{17}$, and $\ket{\Psi_{10}} = \ket{18}$ with an energy value $E_{1-10} = 0$ as well as the molecular wave functions $\ket{\Psi_{11}^\pm} = \frac{1}{\sqrt{2}}\left(\ket{4} \pm \ket{8} \right)$, $\ket{\Psi_{12}^\pm} = \frac{1}{\sqrt{2}}\left(\ket{5} \pm \ket{9} \right)$,  $\ket{\Psi_{13}^\pm} = \frac{1}{\sqrt{2}}\left(\ket{10} \pm \ket{14} \right)$, and $\ket{\Psi_{14}^\pm} = \frac{1}{\sqrt{2}}\left(\ket{11} \pm \ket{15} \right)$ with energies $E_{11-14}^\pm = \pm s^2$.

The strength of the new bond for the weak cluster is given by the matrix element $\bra{\Psi_1}\mathbf{H}\ket{\Psi_{10}}$. As for the medium cluster we obtain no contribution in the first and the second order expansion.
In the third order we have four contributing terms:
\begin{subequations}
\begin{align}
&\bra{\Psi_1}\mathbf{H}_1 \mathbf{P} \frac{1}{E-\mathbf{H}_0}\mathbf{P}\mathbf{H}_1\mathbf{P} \frac{1}{E-\mathbf{H}_0}\mathbf{P}\mathbf{H}_1\ket{\Psi_{10}} \\
 &= \sum \bra{\Psi_1}\mathbf{H}_1 \ket{\Psi_{11}^\pm} \frac{1}{E_1-E_{11}^\pm} \bra{\Psi_{11}^\pm}\mathbf{H}_1\ket{\Psi_{14}^\pm} \nonumber \\ & \quad\;\frac{1}{E_1-E_{14}^\pm} \bra{\Psi_{14}^\pm}\mathbf{H}_1\ket{\Psi_{10}} \\
 &= 4 \frac{w^2}{\sqrt{2}} \frac{w^2}{-2s^2} \frac{w^2}{-\sqrt{2}s^2}
 = \frac{w^6}{s^4} \\
 &\Longrightarrow  \left(w^\prime\right)^2 = \frac{w^6}{s^4} = \frac{w^4}{s^4} w^2
\end{align}
\end{subequations}

Hence, for the atomic RG of the golden-mean labyrinth tiling the scaling of the bond strengths for all three clusters is given by $\bar{z}_\mathrm{Au}^\mathrm{2d} = \left( \bar{z}_\mathrm{Au} \right)^2 = {w^4}/{s^4}$.

\mathversion{bold}
\subsection{Molecular RG}
\mathversion{normal}
\label{subsec:labyrinth-fib-mol}

Also in the molecular RG of the golden-mean labyrinth tiling three types of clusters occur (cf.\ Fig.\ \ref{fig:rg-labyrinth}). The corresponding substitution rules for the strong, medium, and weak clusters are shown in Figs.\ \ref{fig:rg-labyrinth-mol-strong}, \ref{fig:rg-labyrinth-mol-medium}, and \ref{fig:rg-labyrinth-mol-weak}, respectively. Again, the length scaling in two dimensions follows from Eq.\ \eqref{equ:rg-length-scaling-nD} as
\begin{equation}
    c_\mathrm{Au}^\mathrm{mol,2d} = \tau_\mathrm{Au}^{-4} \;.
\end{equation}

\subsubsection{Energy scaling for strong cluster}

\begin{figure}[t!]
    \centering
    \includegraphics[width=0.71\columnwidth]{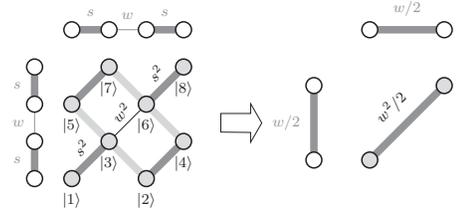}
    \caption{Substitution rule for the strong cluster in the molecular RG approach for the golden-mean labyrinth tiling $\mathcal{L}^\mathrm{Au}$.}
 \label{fig:rg-labyrinth-mol-strong}
\end{figure}

The Hamiltonian of the strong cluster of the molecular RG contains all coupling strengths shown in Fig.\ \ref{fig:rg-labyrinth-mol-strong}, and
the Hamiltonian $\mathbf{H}_0$ yields the four molecular eigenstates
$\ket{\Psi_1^\pm} = \frac{1}{\sqrt{2}}\left(\ket{1} \pm \ket{3} \right)$,
$\ket{\Psi_2^\pm} = \frac{1}{\sqrt{2}}\left(\ket{2} \pm \ket{4} \right)$,
$\ket{\Psi_3^\pm} = \frac{1}{\sqrt{2}}\left(\ket{5} \pm \ket{7} \right)$, and
$\ket{\Psi_4^\pm} = \frac{1}{\sqrt{2}}\left(\ket{6} \pm \ket{8} \right)$
with $E^\pm = \pm s^2$.

The new bond strength for the bonding state is given by the matrix element $\bra{\Psi_1^+}\mathbf{H}\ket{\Psi_4^+}$, where only the leading non-zero term of the perturbation expansion is relevant. The BW perturbation theory already yields a contribution in the first order expansion with $\bra{\Psi_1^+}\mathbf{H}_1\ket{\Psi_4^+} = {w^2}/{2}$. Hence, the scaling of the energies of this cluster is given by $(s^\prime)^2 = \left({w^2}/{2s^2}\right) s^2$.

\subsubsection{Energy scaling for medium cluster}

\begin{figure}[b!]
    \centering
    \includegraphics[width=0.85\columnwidth]{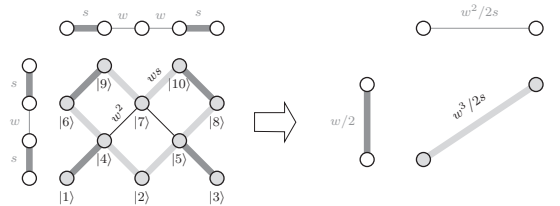}
    \caption{Substitution rule for the medium cluster in the molecular RG for the golden-mean labyrinth tiling $\mathcal{L}^\mathrm{Au}$.}
 \label{fig:rg-labyrinth-mol-medium}
\end{figure}

The Hamiltonian for the medium cluster contains all coupling strengths as shown in Fig.\ \ref{fig:rg-labyrinth-mol-medium}. The eigenstates of $\mathbf{H}_0$ are $\ket{\Psi_1} = \ket{2}$, $\ket{\Psi_2} = \ket{7}$ with $E_{1/2} = 0$ and $\ket{\Psi_{3}^\pm} = \frac{1}{\sqrt{2}}\left(\ket{1} \pm \ket{4} \right)$, $\ket{\Psi_{4}^\pm} = \frac{1}{\sqrt{2}}\left(\ket{3} \pm \ket{5} \right)$, $\ket{\Psi_{5}^\pm} = \frac{1}{\sqrt{2}}\left(\ket{6} \pm \ket{9} \right)$, $\ket{\Psi_{6}^\pm} = \frac{1}{\sqrt{2}}\left(\ket{8} \pm \ket{10} \right)$ with $E_{3-6}^\pm = \pm s^2$.

The strength of the new bond for the bonding state is given by the matrix element $\bra{\Psi_3^+}\mathbf{H}\ket{\Psi_6^+}$. The BW perturbation theory yields no first order contribution because $\bra{\Psi_3^\pm}\mathbf{H}_1\ket{\Psi_6^\pm} = 0$ and the second order contribution for the bonding state is given by
{\allowdisplaybreaks
\begin{subequations}
\begin{align}
 &\bra{\Psi_3^+}\mathbf{H}_1 \mathbf{P} \frac{1}{E-\mathbf{H}_0}\mathbf{P}\mathbf{H}_1\ket{\Psi_6^+} \\
 &= \bra{\Psi_3^+}\mathbf{H}_1 \ket{\Psi_2} \frac{1}{E_3^+-E_2} \bra{\Psi_2}\mathbf{H}_1\ket{\Psi_6^+} \\
 &= \frac{w^2}{\sqrt{2}} \frac{1}{s^2} \frac{ws}{\sqrt{2}} =  \frac{w^3}{2s} \\
 & \Longrightarrow w^\prime s^\prime = \frac{w^3}{2s} = \frac{w^2}{2s^2} ws  \;.
\end{align}
\end{subequations}}

\subsubsection{Energy scaling for weak cluster}

\begin{figure}[b!]
    \centering
    \includegraphics[width=0.85\columnwidth]{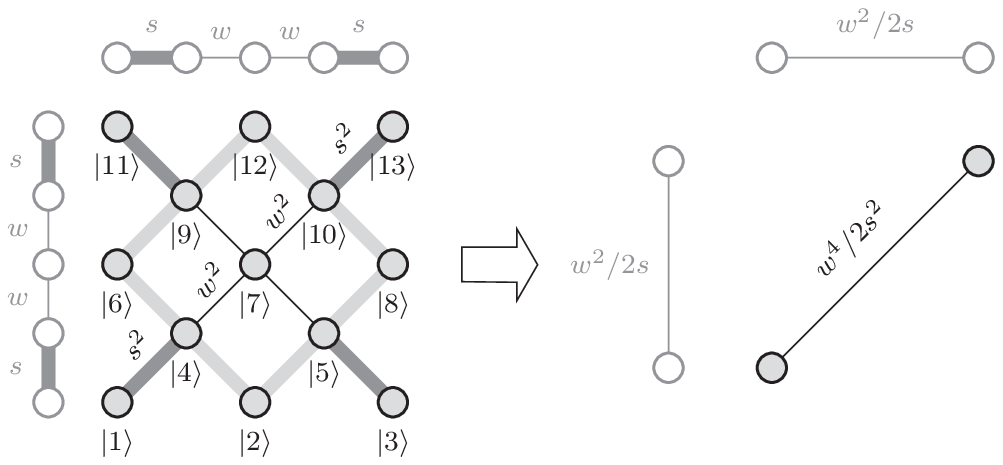}
    \caption{Substitution rule for the weak cluster in the molecular RG approach for the golden-mean labyrinth tiling $\mathcal{L}^\mathrm{Au}$.}
 \label{fig:rg-labyrinth-mol-weak}
\end{figure}

For the weak cluster the Hamiltonian again contains all coupling strengths as shown in Fig.\ \ref{fig:rg-labyrinth-mol-weak}. The Hamiltonian $\mathbf{H}_0$ yields the atomic states $\ket{\Psi_1} = \ket{2}$, $\ket{\Psi_2} = \ket{6}$, $ \ket{\Psi_3} = \ket{7}$, $\ket{\Psi_4} = \ket{8}$ and $\ket{\Psi_5} = \ket{12}$ with $E = 0$ as well as the molecular wave functions $\ket{\Psi_{6}^\pm} = \frac{1}{\sqrt{2}}\left(\ket{1} \pm \ket{4} \right)$, $\ket{\Psi_{7}^\pm} = \frac{1}{\sqrt{2}}\left(\ket{3} \pm \ket{5} \right)$, $\ket{\Psi_{8}^\pm} = \frac{1}{\sqrt{2}}\left(\ket{9} \pm \ket{11} \right)$, and $\ket{\Psi_{9}^\pm} = \frac{1}{\sqrt{2}}\left(\ket{10} \pm \ket{13} \right)$ with the energies $E^\pm = \pm s^2$.

The new bond strength for the bonding state follows from the leading non-zero term of the perturbation expansion of $\bra{\Psi_6^+}\mathbf{H}\ket{\Psi_9^+}$. The BW perturbation theory yields no first order contribution ($\bra{\Psi_6^+}\mathbf{H}_1\ket{\Psi_9^+} = 0$) and the second order contribution is given by
{\allowdisplaybreaks
\begin{subequations}
\begin{align}
 &\bra{\Psi_6^+}\mathbf{H}_1 \mathbf{P} \frac{1}{E-\mathbf{H}_0}\mathbf{P}\mathbf{H}_1\ket{\Psi_9^+} \\
 &= \bra{\Psi_6^+}\mathbf{H}_1 \ket{\Psi_3} \frac{1}{E_6^+-E_3} \bra{\Psi_3}\mathbf{H}_1\ket{\Psi_9^+} \\
 &= \frac{w^2}{\sqrt{2}} \frac{1}{s^2} \frac{w^2}{\sqrt{2}} =  \frac{w^4}{2s^2} \\
 &\Longrightarrow \left(w^\prime\right)^2 = \frac{w^4}{2s^2} = \frac{w^2}{2s^2} w^2  \;.
\end{align}
\end{subequations}}
Hence, the molecular RG yields for all three types of clusters the scaling factor $z_\mathrm{Au}^\mathrm{2d} = {w^2}/{2s^2}$.

\bibliographystyle{apsrev4-1}

\begin{thebibliography}{40}%
\makeatletter
\providecommand \@ifxundefined [1]{%
 \@ifx{#1\undefined}
}%
\providecommand \@ifnum [1]{%
 \ifnum #1\expandafter \@firstoftwo
 \else \expandafter \@secondoftwo
 \fi
}%
\providecommand \@ifx [1]{%
 \ifx #1\expandafter \@firstoftwo
 \else \expandafter \@secondoftwo
 \fi
}%
\providecommand \natexlab [1]{#1}%
\providecommand \enquote  [1]{``#1''}%
\providecommand \bibnamefont  [1]{#1}%
\providecommand \bibfnamefont [1]{#1}%
\providecommand \citenamefont [1]{#1}%
\providecommand \href@noop [0]{\@secondoftwo}%
\providecommand \href [0]{\begingroup \@sanitize@url \@href}%
\providecommand \@href[1]{\@@startlink{#1}\@@href}%
\providecommand \@@href[1]{\endgroup#1\@@endlink}%
\providecommand \@sanitize@url [0]{\catcode `\\12\catcode `\$12\catcode
  `\&12\catcode `\#12\catcode `\^12\catcode `\_12\catcode `\%12\relax}%
\providecommand \@@startlink[1]{}%
\providecommand \@@endlink[0]{}%
\providecommand \url  [0]{\begingroup\@sanitize@url \@url }%
\providecommand \@url [1]{\endgroup\@href {#1}{\urlprefix }}%
\providecommand \urlprefix  [0]{URL }%
\providecommand \Eprint [0]{\href }%
\providecommand \doibase [0]{http://dx.doi.org/}%
\providecommand \selectlanguage [0]{\@gobble}%
\providecommand \bibinfo  [0]{\@secondoftwo}%
\providecommand \bibfield  [0]{\@secondoftwo}%
\providecommand \translation [1]{[#1]}%
\providecommand \BibitemOpen [0]{}%
\providecommand \bibitemStop [0]{}%
\providecommand \bibitemNoStop [0]{.\EOS\space}%
\providecommand \EOS [0]{\spacefactor3000\relax}%
\providecommand \BibitemShut  [1]{\csname bibitem#1\endcsname}%
\let\auto@bib@innerbib\@empty
\bibitem [{\citenamefont {{S}hechtman}\ \emph {et~al.}(1984)\citenamefont
  {{S}hechtman}, \citenamefont {{B}lech}, \citenamefont {{G}ratias},\ and\
  \citenamefont {{C}ahn}}]{PhysRevLett.1984.Shechtman}%
  \BibitemOpen
  \bibfield  {author} {\bibinfo {author} {\bibfnamefont {D.}~\bibnamefont
  {{S}hechtman}}, \bibinfo {author} {\bibfnamefont {I.}~\bibnamefont
  {{B}lech}}, \bibinfo {author} {\bibfnamefont {D.}~\bibnamefont {{G}ratias}},
  \ and\ \bibinfo {author} {\bibfnamefont {J.~W.}\ \bibnamefont {{C}ahn}},\
  }\href {\doibase 10.1103/PhysRevLett.53.1951} {\bibfield  {journal} {\bibinfo
   {journal} {{P}hys. {R}ev. {L}ett.}\ }\textbf {\bibinfo {volume} {53}},\
  \bibinfo {pages} {1951} (\bibinfo {year} {1984})}\BibitemShut {NoStop}%
\bibitem [{\citenamefont {{A}mmann}\ \emph {et~al.}(1992)\citenamefont
  {{A}mmann}, \citenamefont {{G}r\"unbaum},\ and\ \citenamefont
  {{S}hephard}}]{DCGeom.1992.Ammann}%
  \BibitemOpen
  \bibfield  {author} {\bibinfo {author} {\bibfnamefont {R.}~\bibnamefont
  {{A}mmann}}, \bibinfo {author} {\bibfnamefont {B.}~\bibnamefont
  {{G}r\"unbaum}}, \ and\ \bibinfo {author} {\bibfnamefont {G.~C.}\
  \bibnamefont {{S}hephard}},\ }\href {\doibase 10.1007/BF02293033} {\bibfield
  {journal} {\bibinfo  {journal} {{D}iscrete \& {C}omputational {G}eometry}\
  }\textbf {\bibinfo {volume} {8}},\ \bibinfo {pages} {1} (\bibinfo {year}
  {1992})}\BibitemShut {NoStop}%
\bibitem [{\citenamefont {{P}enrose}(1974)}]{BIMathApp.1974.Penrose}%
  \BibitemOpen
  \bibfield  {author} {\bibinfo {author} {\bibfnamefont {R.}~\bibnamefont
  {{P}enrose}},\ }\href@noop {} {\bibfield  {journal} {\bibinfo  {journal}
  {{B}ull. {I}nst. {M}ath. {A}ppl.}\ }\textbf {\bibinfo {volume} {10}},\
  \bibinfo {pages} {266} (\bibinfo {year} {1974})}\BibitemShut {NoStop}%
\bibitem [{\citenamefont {{G}ardner}(1977)}]{SciAm.1977.Gardner}%
  \BibitemOpen
  \bibfield  {author} {\bibinfo {author} {\bibfnamefont {M.}~\bibnamefont
  {{G}ardner}},\ }\href {\doibase 10.1038/scientificamerican0177-110}
  {\bibfield  {journal} {\bibinfo  {journal} {{S}cientific {A}merican}\
  }\textbf {\bibinfo {volume} {236}},\ \bibinfo {pages} {110} (\bibinfo {year}
  {1977})}\BibitemShut {NoStop}%
\bibitem [{\citenamefont {{D}ubois}(2005)}]{UsefulQuasicrystals}%
  \BibitemOpen
  \bibfield  {author} {\bibinfo {author} {\bibfnamefont {J.-M.}\ \bibnamefont
  {{D}ubois}},\ }\href@noop {} {\emph {\bibinfo {title} {{U}seful
  {Q}uasicrystals}}}\ (\bibinfo  {publisher} {World Scientific},\ \bibinfo
  {address} {Singapore London},\ \bibinfo {year} {2005})\BibitemShut {NoStop}%
\bibitem [{\citenamefont {{S}tadnik}(1999)}]{PhysicalProperties.1999.Stadnik}%
  \BibitemOpen
  \bibfield  {author} {\bibinfo {author} {\bibfnamefont {Z.~M.}\ \bibnamefont
  {{S}tadnik}},\ }\href@noop {} {\emph {\bibinfo {title} {{P}hysical
  {P}roperties of {Q}uasicrystals}}}\ (\bibinfo  {publisher} {Springer},\
  \bibinfo {address} {Berlin Heidelberg New York},\ \bibinfo {year}
  {1999})\BibitemShut {NoStop}%
\bibitem [{\citenamefont {{R}oche}\ \emph {et~al.}(1997)\citenamefont
  {{R}oche}, \citenamefont {{T}rambly~de {L}aissardi\`{e}re},\ and\
  \citenamefont {{M}ayou}}]{JMathPhys.1997.Roche}%
  \BibitemOpen
  \bibfield  {author} {\bibinfo {author} {\bibfnamefont {S.}~\bibnamefont
  {{R}oche}}, \bibinfo {author} {\bibfnamefont {G.}~\bibnamefont {{T}rambly~de
  {L}aissardi\`{e}re}}, \ and\ \bibinfo {author} {\bibfnamefont
  {D.}~\bibnamefont {{M}ayou}},\ }\href {\doibase 10.1063/1.531914} {\bibfield
  {journal} {\bibinfo  {journal} {{J}. {M}ath. {P}hys.}\ }\textbf {\bibinfo
  {volume} {38}},\ \bibinfo {pages} {1794} (\bibinfo {year}
  {1997})}\BibitemShut {NoStop}%
\bibitem [{\citenamefont {{M}ayou}\ \emph {et~al.}(1993)\citenamefont
  {{M}ayou}, \citenamefont {{B}erger}, \citenamefont {{Cyrot-Lackmann}},
  \citenamefont {{K}lein},\ and\ \citenamefont
  {{L}anco}}]{PhysRevLett.1993.Mayou}%
  \BibitemOpen
  \bibfield  {author} {\bibinfo {author} {\bibfnamefont {D.}~\bibnamefont
  {{M}ayou}}, \bibinfo {author} {\bibfnamefont {C.}~\bibnamefont {{B}erger}},
  \bibinfo {author} {\bibfnamefont {F.}~\bibnamefont {{Cyrot-Lackmann}}},
  \bibinfo {author} {\bibfnamefont {T.}~\bibnamefont {{K}lein}}, \ and\
  \bibinfo {author} {\bibfnamefont {P.}~\bibnamefont {{L}anco}},\ }\href
  {\doibase 10.1103/PhysRevLett.70.3915} {\bibfield  {journal} {\bibinfo
  {journal} {{P}hys. {R}ev. {L}ett.}\ }\textbf {\bibinfo {volume} {70}},\
  \bibinfo {pages} {3915} (\bibinfo {year} {1993})}\BibitemShut {NoStop}%
\bibitem [{\citenamefont {{P}oon}(1992)}]{AdvPhys.1992.Poon}%
  \BibitemOpen
  \bibfield  {author} {\bibinfo {author} {\bibfnamefont {S.~J.}\ \bibnamefont
  {{P}oon}},\ }\href {\doibase 0.1080/00018739200101513} {\bibfield  {journal}
  {\bibinfo  {journal} {{A}dv. {P}hys.}\ }\textbf {\bibinfo {volume} {41}},\
  \bibinfo {pages} {303} (\bibinfo {year} {1992})}\BibitemShut {NoStop}%
\bibitem [{\citenamefont {{D}amanik}(2000)}]{MathQuasi.2000.Damanik}%
  \BibitemOpen
  \bibfield  {author} {\bibinfo {author} {\bibfnamefont {D.}~\bibnamefont
  {{D}amanik}},\ }in\ \href@noop {} {\emph {\bibinfo {booktitle} {Directions in
  Mathematical Quasicrystals}}},\ Vol.~\bibinfo {volume} {13},\ \bibinfo
  {editor} {edited by\ \bibinfo {editor} {\bibfnamefont {M.}~\bibnamefont
  {Baake}}\ and\ \bibinfo {editor} {\bibfnamefont {R.~V.}\ \bibnamefont
  {Moody}}}\ (\bibinfo  {publisher} {AMS, Providence},\ \bibinfo {year}
  {2000})\ pp.\ \bibinfo {pages} {277--304}\BibitemShut {NoStop}%
\bibitem [{\citenamefont {{K}ohmoto}\ \emph {et~al.}(1987)\citenamefont
  {{K}ohmoto}, \citenamefont {{S}utherland},\ and\ \citenamefont
  {{T}ang}}]{PhysRevB.1987.Kohmoto}%
  \BibitemOpen
  \bibfield  {author} {\bibinfo {author} {\bibfnamefont {M.}~\bibnamefont
  {{K}ohmoto}}, \bibinfo {author} {\bibfnamefont {B.}~\bibnamefont
  {{S}utherland}}, \ and\ \bibinfo {author} {\bibfnamefont {C.}~\bibnamefont
  {{T}ang}},\ }\href {\doibase 10.1103/PhysRevB.35.1020} {\bibfield  {journal}
  {\bibinfo  {journal} {{P}hys. {R}ev. {B}}\ }\textbf {\bibinfo {volume}
  {35}},\ \bibinfo {pages} {1020} (\bibinfo {year} {1987})}\BibitemShut
  {NoStop}%
\bibitem [{\citenamefont {{S}ire}\ and\ \citenamefont
  {{M}osseri}(1989)}]{JPhysFrance.1989.Sire}%
  \BibitemOpen
  \bibfield  {author} {\bibinfo {author} {\bibfnamefont {C.}~\bibnamefont
  {{S}ire}}\ and\ \bibinfo {author} {\bibfnamefont {R.}~\bibnamefont
  {{M}osseri}},\ }\href {\doibase 10.1051/jphys:0198900500240344700} {\bibfield
   {journal} {\bibinfo  {journal} {{J}. {P}hys. {F}rance}\ }\textbf {\bibinfo
  {volume} {50}},\ \bibinfo {pages} {3447} (\bibinfo {year}
  {1989})}\BibitemShut {NoStop}%
\bibitem [{\citenamefont {{G}rimm}\ and\ \citenamefont
  {{S}chreiber}(2003)}]{Quasicrystals.2003.Grimm}%
  \BibitemOpen
  \bibfield  {author} {\bibinfo {author} {\bibfnamefont {U.}~\bibnamefont
  {{G}rimm}}\ and\ \bibinfo {author} {\bibfnamefont {M.}~\bibnamefont
  {{S}chreiber}},\ }in\ \href@noop {} {\emph {\bibinfo {booktitle}
  {Quasicrystals: Structure and Physical Properties}}},\ \bibinfo {editor}
  {edited by\ \bibinfo {editor} {\bibfnamefont {H.-R.}\ \bibnamefont
  {Trebin}}}\ (\bibinfo  {publisher} {Wiley-VCH, Berlin},\ \bibinfo {year}
  {2003})\ pp.\ \bibinfo {pages} {210--235}\BibitemShut {NoStop}%
\bibitem [{\citenamefont {{P}assaro}\ \emph {et~al.}(1992)\citenamefont
  {{P}assaro}, \citenamefont {{S}ire},\ and\ \citenamefont
  {{B}enza}}]{PhysRevB.1992.Passaro}%
  \BibitemOpen
  \bibfield  {author} {\bibinfo {author} {\bibfnamefont {B.}~\bibnamefont
  {{P}assaro}}, \bibinfo {author} {\bibfnamefont {C.}~\bibnamefont {{S}ire}}, \
  and\ \bibinfo {author} {\bibfnamefont {V.~G.}\ \bibnamefont {{B}enza}},\
  }\href {\doibase 10.1103/PhysRevB.46.13751} {\bibfield  {journal} {\bibinfo
  {journal} {{P}hys. {R}ev. {B}}\ }\textbf {\bibinfo {volume} {46}},\ \bibinfo
  {pages} {13751} (\bibinfo {year} {1992})}\BibitemShut {NoStop}%
\bibitem [{\citenamefont {{T}riozon}\ \emph {et~al.}(2002)\citenamefont
  {{T}riozon}, \citenamefont {{V}idal}, \citenamefont {{M}osseri},\ and\
  \citenamefont {{M}ayou}}]{PhysRevB.2002.Triozon}%
  \BibitemOpen
  \bibfield  {author} {\bibinfo {author} {\bibfnamefont {F.}~\bibnamefont
  {{T}riozon}}, \bibinfo {author} {\bibfnamefont {J.}~\bibnamefont {{V}idal}},
  \bibinfo {author} {\bibfnamefont {R.}~\bibnamefont {{M}osseri}}, \ and\
  \bibinfo {author} {\bibfnamefont {D.}~\bibnamefont {{M}ayou}},\ }\href
  {\doibase 10.1103/PhysRevB.65.220202} {\bibfield  {journal} {\bibinfo
  {journal} {{P}hys. {R}ev. {B}}\ }\textbf {\bibinfo {volume} {65}},\ \bibinfo
  {pages} {220202} (\bibinfo {year} {2002})}\BibitemShut {NoStop}%
\bibitem [{\citenamefont {{V}idal}\ \emph {et~al.}(2003)\citenamefont
  {{V}idal}, \citenamefont {{D}estainville},\ and\ \citenamefont
  {{M}osseri}}]{PhysRevB.2003.Vidal}%
  \BibitemOpen
  \bibfield  {author} {\bibinfo {author} {\bibfnamefont {J.}~\bibnamefont
  {{V}idal}}, \bibinfo {author} {\bibfnamefont {N.}~\bibnamefont
  {{D}estainville}}, \ and\ \bibinfo {author} {\bibfnamefont {R.}~\bibnamefont
  {{M}osseri}},\ }\href {\doibase 10.1103/PhysRevB.68.172202} {\bibfield
  {journal} {\bibinfo  {journal} {{P}hys. {R}ev. {B}}\ }\textbf {\bibinfo
  {volume} {68}},\ \bibinfo {pages} {172202} (\bibinfo {year}
  {2003})}\BibitemShut {NoStop}%
\bibitem [{\citenamefont {{R}ieth}\ and\ \citenamefont
  {{S}chreiber}(1995)}]{PhysRevB.1995.Rieth}%
  \BibitemOpen
  \bibfield  {author} {\bibinfo {author} {\bibfnamefont {T.}~\bibnamefont
  {{R}ieth}}\ and\ \bibinfo {author} {\bibfnamefont {M.}~\bibnamefont
  {{S}chreiber}},\ }\href {\doibase 10.1103/PhysRevB.51.15827} {\bibfield
  {journal} {\bibinfo  {journal} {{P}hys. {R}ev. {B}}\ }\textbf {\bibinfo
  {volume} {51}},\ \bibinfo {pages} {15827} (\bibinfo {year}
  {1995})}\BibitemShut {NoStop}%
\bibitem [{\citenamefont {{R}epetowicz}\ \emph {et~al.}(1998)\citenamefont
  {{R}epetowicz}, \citenamefont {{G}rimm},\ and\ \citenamefont
  {{S}chreiber}}]{PhysRevB.1998.Repetowicz}%
  \BibitemOpen
  \bibfield  {author} {\bibinfo {author} {\bibfnamefont {P.}~\bibnamefont
  {{R}epetowicz}}, \bibinfo {author} {\bibfnamefont {U.}~\bibnamefont
  {{G}rimm}}, \ and\ \bibinfo {author} {\bibfnamefont {M.}~\bibnamefont
  {{S}chreiber}},\ }\href {\doibase 10.1103/PhysRevB.58.13482} {\bibfield
  {journal} {\bibinfo  {journal} {{P}hys. {R}ev. {B}}\ }\textbf {\bibinfo
  {volume} {58}},\ \bibinfo {pages} {13482} (\bibinfo {year}
  {1998})}\BibitemShut {NoStop}%
\bibitem [{\citenamefont {{S}ire}(1989)}]{EurophysLett.1989.Sire}%
  \BibitemOpen
  \bibfield  {author} {\bibinfo {author} {\bibfnamefont {C.}~\bibnamefont
  {{S}ire}},\ }\href {\doibase 10.1209/0295-5075/10/5/016} {\bibfield
  {journal} {\bibinfo  {journal} {{E}urophys. {L}ett.}\ }\textbf {\bibinfo
  {volume} {10}},\ \bibinfo {pages} {483} (\bibinfo {year} {1989})}\BibitemShut
  {NoStop}%
\bibitem [{\citenamefont {{S}ire}\ \emph {et~al.}(1989)\citenamefont {{S}ire},
  \citenamefont {{M}osseri},\ and\ \citenamefont
  {{S}adoc}}]{JPhysFrance.1989.Sire2}%
  \BibitemOpen
  \bibfield  {author} {\bibinfo {author} {\bibfnamefont {C.}~\bibnamefont
  {{S}ire}}, \bibinfo {author} {\bibfnamefont {R.}~\bibnamefont {{M}osseri}}, \
  and\ \bibinfo {author} {\bibfnamefont {J.-F.}\ \bibnamefont {{S}adoc}},\
  }\href {\doibase 10.1051/jphys:0198900500240346300} {\bibfield  {journal}
  {\bibinfo  {journal} {{J}. {P}hys. {F}rance}\ }\textbf {\bibinfo {volume}
  {50}},\ \bibinfo {pages} {3463} (\bibinfo {year} {1989})}\BibitemShut
  {NoStop}%
\bibitem [{\citenamefont {{A}be}\ and\ \citenamefont
  {{H}iramoto}(1987)}]{PhysRevA.1987.Abe}%
  \BibitemOpen
  \bibfield  {author} {\bibinfo {author} {\bibfnamefont {S.}~\bibnamefont
  {{A}be}}\ and\ \bibinfo {author} {\bibfnamefont {H.}~\bibnamefont
  {{H}iramoto}},\ }\href {\doibase 10.1103/PhysRevA.36.5349} {\bibfield
  {journal} {\bibinfo  {journal} {{P}hys. {R}ev. {A}}\ }\textbf {\bibinfo
  {volume} {36}},\ \bibinfo {pages} {5349} (\bibinfo {year}
  {1987})}\BibitemShut {NoStop}%
\bibitem [{\citenamefont {{H}iramoto}\ and\ \citenamefont
  {{A}be}(1988)}]{JPhysJap.1988.Hiramoto}%
  \BibitemOpen
  \bibfield  {author} {\bibinfo {author} {\bibfnamefont {H.}~\bibnamefont
  {{H}iramoto}}\ and\ \bibinfo {author} {\bibfnamefont {S.}~\bibnamefont
  {{A}be}},\ }\href {\doibase 10.1143/JPSJ.57.230} {\bibfield  {journal}
  {\bibinfo  {journal} {{J}. {P}hys. {S}oc. {J}pn.}\ }\textbf {\bibinfo
  {volume} {57}},\ \bibinfo {pages} {230} (\bibinfo {year} {1988})}\BibitemShut
  {NoStop}%
\bibitem [{\citenamefont {{P}i\'echon}(1996)}]{PhysRevLett.1996.Piechon}%
  \BibitemOpen
  \bibfield  {author} {\bibinfo {author} {\bibfnamefont {F.}~\bibnamefont
  {{P}i\'echon}},\ }\href {\doibase 10.1103/PhysRevLett.76.4372} {\bibfield
  {journal} {\bibinfo  {journal} {{P}hys. {R}ev. {L}ett.}\ }\textbf {\bibinfo
  {volume} {76}},\ \bibinfo {pages} {4372} (\bibinfo {year}
  {1996})}\BibitemShut {NoStop}%
\bibitem [{\citenamefont {{T}hiem}\ \emph {et~al.}(2009)\citenamefont
  {{T}hiem}, \citenamefont {{S}chreiber},\ and\ \citenamefont
  {{G}rimm}}]{PhysRevB.2009.Thiem}%
  \BibitemOpen
  \bibfield  {author} {\bibinfo {author} {\bibfnamefont {S.}~\bibnamefont
  {{T}hiem}}, \bibinfo {author} {\bibfnamefont {M.}~\bibnamefont
  {{S}chreiber}}, \ and\ \bibinfo {author} {\bibfnamefont {U.}~\bibnamefont
  {{G}rimm}},\ }\href {\doibase 10.1103/PhysRevB.80.214203} {\bibfield
  {journal} {\bibinfo  {journal} {{P}hys. {R}ev. {B}}\ }\textbf {\bibinfo
  {volume} {80}},\ \bibinfo {pages} {214203} (\bibinfo {year}
  {2009})}\BibitemShut {NoStop}%
\bibitem [{\citenamefont {{G}umbs}\ and\ \citenamefont
  {{A}li}(1989)}]{JPhysA.1989.Gumbs}%
  \BibitemOpen
  \bibfield  {author} {\bibinfo {author} {\bibfnamefont {G.}~\bibnamefont
  {{G}umbs}}\ and\ \bibinfo {author} {\bibfnamefont {M.~K.}\ \bibnamefont
  {{A}li}},\ }\href {\doibase 10.1088/0305-4470/22/8/012} {\bibfield  {journal}
  {\bibinfo  {journal} {{J}. {P}hys. {A}}\ }\textbf {\bibinfo {volume} {22}},\
  \bibinfo {pages} {951} (\bibinfo {year} {1989})}\BibitemShut {NoStop}%
\bibitem [{\citenamefont {{S}ocolar}\ and\ \citenamefont
  {{S}teinhardt}(1986)}]{PhysRevB.1986.Socolar}%
  \BibitemOpen
  \bibfield  {author} {\bibinfo {author} {\bibfnamefont {J.~E.~S.}\
  \bibnamefont {{S}ocolar}}\ and\ \bibinfo {author} {\bibfnamefont {P.~J.}\
  \bibnamefont {{S}teinhardt}},\ }\href {\doibase 10.1103/PhysRevB.34.617}
  {\bibfield  {journal} {\bibinfo  {journal} {{P}hys. {R}ev. {B}}\ }\textbf
  {\bibinfo {volume} {34}},\ \bibinfo {pages} {617} (\bibinfo {year}
  {1986})}\BibitemShut {NoStop}%
\bibitem [{\citenamefont {{Y}uan}\ \emph {et~al.}(2000)\citenamefont {{Y}uan},
  \citenamefont {{G}rimm}, \citenamefont {{R}epetowicz},\ and\ \citenamefont
  {{S}chreiber}}]{PhysRevB.2000.Yuan}%
  \BibitemOpen
  \bibfield  {author} {\bibinfo {author} {\bibfnamefont {H.~Q.}\ \bibnamefont
  {{Y}uan}}, \bibinfo {author} {\bibfnamefont {U.}~\bibnamefont {{G}rimm}},
  \bibinfo {author} {\bibfnamefont {P.}~\bibnamefont {{R}epetowicz}}, \ and\
  \bibinfo {author} {\bibfnamefont {M.}~\bibnamefont {{S}chreiber}},\ }\href
  {\doibase 10.1103/PhysRevB.62.15569} {\bibfield  {journal} {\bibinfo
  {journal} {{P}hys. {R}ev. {B}}\ }\textbf {\bibinfo {volume} {62}},\ \bibinfo
  {pages} {15569} (\bibinfo {year} {2000})}\BibitemShut {NoStop}%
\bibitem [{\citenamefont {{C}erovski}\ \emph {et~al.}(2005)\citenamefont
  {{C}erovski}, \citenamefont {{S}chreiber},\ and\ \citenamefont
  {{G}rimm}}]{PhysRevB.2005.Cerovski}%
  \BibitemOpen
  \bibfield  {author} {\bibinfo {author} {\bibfnamefont {V.~Z.}\ \bibnamefont
  {{C}erovski}}, \bibinfo {author} {\bibfnamefont {M.}~\bibnamefont
  {{S}chreiber}}, \ and\ \bibinfo {author} {\bibfnamefont {U.}~\bibnamefont
  {{G}rimm}},\ }\href {\doibase 10.1103/PhysRevB.72.054203} {\bibfield
  {journal} {\bibinfo  {journal} {{P}hys. {R}ev. {B}}\ }\textbf {\bibinfo
  {volume} {72}},\ \bibinfo {eid} {054203} (\bibinfo {year}
  {2005})}\BibitemShut {NoStop}%
\bibitem [{\citenamefont {{T}hiem}\ and\ \citenamefont
  {{S}chreiber}(2010)}]{JPhysCS.2010.Thiem}%
  \BibitemOpen
  \bibfield  {author} {\bibinfo {author} {\bibfnamefont {S.}~\bibnamefont
  {{T}hiem}}\ and\ \bibinfo {author} {\bibfnamefont {M.}~\bibnamefont
  {{S}chreiber}},\ }\href {\doibase 10.1088/1742-6596/226/1/012029} {\bibfield
  {journal} {\bibinfo  {journal} {{J}. {P}hys.: {C}onf. {S}er.}\ }\textbf
  {\bibinfo {volume} {226}},\ \bibinfo {pages} {012029} (\bibinfo {year}
  {2010})}\BibitemShut {NoStop}%
\bibitem [{\citenamefont {{Z}hong}\ and\ \citenamefont
  {{M}osseri}(1995)}]{JPhys.1995.Zhong}%
  \BibitemOpen
  \bibfield  {author} {\bibinfo {author} {\bibfnamefont {J.}~\bibnamefont
  {{Z}hong}}\ and\ \bibinfo {author} {\bibfnamefont {R.}~\bibnamefont
  {{M}osseri}},\ }\href {\doibase 10.1088/0953-8984/7/44/008} {\bibfield
  {journal} {\bibinfo  {journal} {{J}. {P}hys.: {C}ond. {M}att.}\ }\textbf
  {\bibinfo {volume} {7}},\ \bibinfo {pages} {8383} (\bibinfo {year}
  {1995})}\BibitemShut {NoStop}%
\bibitem [{\citenamefont {{Schulz-Baldes}}\ and\ \citenamefont
  {{B}ellissard}(1998)}]{RevMathPhys.1998.SchulzBaldes}%
  \BibitemOpen
  \bibfield  {author} {\bibinfo {author} {\bibfnamefont {H.}~\bibnamefont
  {{Schulz-Baldes}}}\ and\ \bibinfo {author} {\bibfnamefont {J.}~\bibnamefont
  {{B}ellissard}},\ }\href {\doibase 10.1142/S0129055X98000021} {\bibfield
  {journal} {\bibinfo  {journal} {{R}ev. {M}ath. {P}hys.}\ }\textbf {\bibinfo
  {volume} {10}},\ \bibinfo {pages} {1} (\bibinfo {year} {1998})}\BibitemShut
  {NoStop}%
\bibitem [{\citenamefont {{H}uckestein}\ and\ \citenamefont
  {{S}chweitzer}(1994)}]{PhysRevLett.1994.Huckestein}%
  \BibitemOpen
  \bibfield  {author} {\bibinfo {author} {\bibfnamefont {B.}~\bibnamefont
  {{H}uckestein}}\ and\ \bibinfo {author} {\bibfnamefont {L.}~\bibnamefont
  {{S}chweitzer}},\ }\href {\doibase 10.1103/PhysRevLett.72.713} {\bibfield
  {journal} {\bibinfo  {journal} {{P}hys. {R}ev. {L}ett.}\ }\textbf {\bibinfo
  {volume} {72}},\ \bibinfo {pages} {713} (\bibinfo {year} {1994})}\BibitemShut
  {NoStop}%
\bibitem [{\citenamefont {{M}ayou}(2000)}]{PhysRevLett.2000.Mayou}%
  \BibitemOpen
  \bibfield  {author} {\bibinfo {author} {\bibfnamefont {D.}~\bibnamefont
  {{M}ayou}},\ }\href {\doibase 10.1103/PhysRevLett.85.1290} {\bibfield
  {journal} {\bibinfo  {journal} {{P}hys. {R}ev. {L}ett.}\ }\textbf {\bibinfo
  {volume} {85}},\ \bibinfo {pages} {1290} (\bibinfo {year}
  {2000})}\BibitemShut {NoStop}%
\bibitem [{\citenamefont
  {{Schulz-Baldes}}(1997)}]{PhysRevLett.1997.Schulz-Baldes}%
  \BibitemOpen
  \bibfield  {author} {\bibinfo {author} {\bibfnamefont {H.}~\bibnamefont
  {{Schulz-Baldes}}},\ }\href {\doibase 10.1103/PhysRevLett.78.2176} {\bibfield
   {journal} {\bibinfo  {journal} {{P}hys. {R}ev. {L}ett.}\ }\textbf {\bibinfo
  {volume} {78}},\ \bibinfo {pages} {2176} (\bibinfo {year}
  {1997})}\BibitemShut {NoStop}%
\bibitem [{\citenamefont {{N}iu}\ and\ \citenamefont
  {{N}ori}(1986)}]{PhysRevLett.1986.Niu}%
  \BibitemOpen
  \bibfield  {author} {\bibinfo {author} {\bibfnamefont {Q.}~\bibnamefont
  {{N}iu}}\ and\ \bibinfo {author} {\bibfnamefont {F.}~\bibnamefont {{N}ori}},\
  }\href {\doibase 10.1103/PhysRevLett.57.2057} {\bibfield  {journal} {\bibinfo
   {journal} {{P}hys. {R}ev. {L}ett.}\ }\textbf {\bibinfo {volume} {57}},\
  \bibinfo {pages} {2057} (\bibinfo {year} {1986})}\BibitemShut {NoStop}%
\bibitem [{\citenamefont {{N}iu}\ and\ \citenamefont
  {{N}ori}(1990)}]{PhysRevB.1990.Niu}%
  \BibitemOpen
  \bibfield  {author} {\bibinfo {author} {\bibfnamefont {Q.}~\bibnamefont
  {{N}iu}}\ and\ \bibinfo {author} {\bibfnamefont {F.}~\bibnamefont {{N}ori}},\
  }\href {\doibase 10.1103/PhysRevB.42.10329} {\bibfield  {journal} {\bibinfo
  {journal} {{P}hys. {R}ev. {B}}\ }\textbf {\bibinfo {volume} {42}},\ \bibinfo
  {pages} {10329} (\bibinfo {year} {1990})}\BibitemShut {NoStop}%
\bibitem [{\citenamefont {{W}ilkinson}\ and\ \citenamefont
  {{A}ustin}(1994)}]{PhysRevB.1994.Wilkinson}%
  \BibitemOpen
  \bibfield  {author} {\bibinfo {author} {\bibfnamefont {M.}~\bibnamefont
  {{W}ilkinson}}\ and\ \bibinfo {author} {\bibfnamefont {E.~J.}\ \bibnamefont
  {{A}ustin}},\ }\href {\doibase 10.1103/PhysRevB.50.1420} {\bibfield
  {journal} {\bibinfo  {journal} {{P}hys. {R}ev. {B}}\ }\textbf {\bibinfo
  {volume} {50}},\ \bibinfo {pages} {1420} (\bibinfo {year}
  {1994})}\BibitemShut {NoStop}%
\bibitem [{\citenamefont {{D}amanik}(2006)}]{PhilMag.2006.Damanik}%
  \BibitemOpen
  \bibfield  {author} {\bibinfo {author} {\bibfnamefont {D.}~\bibnamefont
  {{D}amanik}},\ }\href {\doibase 10.1080/14786430500199286} {\bibfield
  {journal} {\bibinfo  {journal} {{P}hilos. {M}ag.}\ }\textbf {\bibinfo
  {volume} {86}},\ \bibinfo {pages} {883} (\bibinfo {year} {2006})}\BibitemShut
  {NoStop}%
\bibitem [{\citenamefont {{D}amanik}\ and\ \citenamefont
  {{T}cheremchantsev}(2007)}]{JAmMath.2007.Damanik}%
  \BibitemOpen
  \bibfield  {author} {\bibinfo {author} {\bibfnamefont {D.}~\bibnamefont
  {{D}amanik}}\ and\ \bibinfo {author} {\bibfnamefont {S.}~\bibnamefont
  {{T}cheremchantsev}},\ }\href {\doibase 10.1090/S0894-0347-06-00554-6}
  {\bibfield  {journal} {\bibinfo  {journal} {{J}. {A}mer. {M}ath. {S}oc.}\
  }\textbf {\bibinfo {volume} {20}},\ \bibinfo {pages} {799} (\bibinfo {year}
  {2007})}\BibitemShut {NoStop}%
\bibitem [{\citenamefont {{B}aake}\ \emph {et~al.}(1993)\citenamefont
  {{B}aake}, \citenamefont {{G}rimm},\ and\ \citenamefont
  {{J}oseph}}]{ModPhys.1993.Baake}%
  \BibitemOpen
  \bibfield  {author} {\bibinfo {author} {\bibfnamefont {M.}~\bibnamefont
  {{B}aake}}, \bibinfo {author} {\bibfnamefont {U.}~\bibnamefont {{G}rimm}}, \
  and\ \bibinfo {author} {\bibfnamefont {D.}~\bibnamefont {{J}oseph}},\ }\href
  {\doibase 10.1142/S021797929300247X} {\bibfield  {journal} {\bibinfo
  {journal} {{I}nt. {J}. {M}od. {P}hys. {B}}\ }\textbf {\bibinfo {volume}
  {7}},\ \bibinfo {pages} {1527} (\bibinfo {year} {1993})}\BibitemShut
  {NoStop}%
\end{thebibliography}

\newcommand{\noopsort}[1]{} \newcommand{\printfirst}[2]{#1}
\newcommand{\singleletter}[1]{#1} \newcommand{\switchargs}[2]{#2#1}

\end{document}